%% file: dust.tex
\documentclass[fleqn,usenatbib]{mnras}

\usepackage{newtxtext,newtxmath}

\usepackage[T1]{fontenc}
\usepackage{ae,aecompl}

\usepackage{graphicx}	
\usepackage{amsmath}	
\usepackage{amssymb}	
\usepackage{etoolbox}
\makeatletter
\patchcmd\@combinedblfloats{\box\@outputbox}{\unvbox\@outputbox}{}{%
	\errmessage{\noexpand\@combinedblfloats could not be patched}%
}%
\makeatother
\usepackage{color} 
\usepackage[normalem]{ulem} 
\usepackage{hyperref} 

\newcommand{\dd}{\mathrm{d}}
\newcommand{\lgal}{{\sc L-Galaxies}}
\newcommand{\Msun}{\mbox{M$_\odot$}}
\newcommand{\tOneA}{{type\,Ia}}
\newcommand{\TOneA}{{Type\,Ia}}
\newcommand{\tTwo}{{type\,II}}

\newenvironment{shortitem}
{\begin{list}{$\bullet$}{\topsep=0pt\itemsep=0pt\parsep=0pt\parskip=0pt\leftmargin=12pt}}
{\end{list}}

\newcommand{\eg}[0]{$\textnormal{e.g. }$}
\newcommand{\ie}[0]{$\textnormal{i.e. }$}
\newcommand{\tn}[1]{\textnormal{#1}}
\newcommand{\sub}[1]{_{\textnormal{#1}}}

\newcommand{\Eq}[1]{Equation~\ref{eq:#1}}
\newcommand{\Fig}[1]{Figure~\ref{fig:#1}}
\newcommand{\Sec}[1]{Section~\ref{sec:#1}}
\newcommand{\Tab}[1]{Table~\ref{tab:#1}}

\defcitealias{Henriques2015}{HWT15}
\defcitealias{Popping2017}{PSG17}
\defcitealias{Zhukovska2008}{ZGT08}

\title[Dust properties in \lgal]{Detailed dust modelling in the \lgal\ semi-analytic model of galaxy formation}

\author[Vijayan et al.]
{Aswin P. Vijayan$^{1}$\thanks{E-mail: A.Payyoor-Vijayan@sussex.ac.uk},
Scott J. Clay$^{1}$,
Peter A. Thomas$^{1}$,
Robert M. Yates$^{2}$,
\newauthor 
Stephen M. Wilkins$^{1}$,
Bruno M. Henriques$^{3}$
\\
$^{1}$Astronomy Centre, University of Sussex, Falmer, Brighton BN1 9QH, UK\\
$^{2}$Max-Planck-Institut f\"ur Astrophysik, Karl-Schwarzschild-Str. 1, D-85741 Garching b. M\"unchen, Germany\\
$^{3}$Institute for Astronomy, ETH Zurich, CH-8093 Zurich, Switzerland
}

\date{Accepted XXX. Received YYY; in original form ZZZ}

\pubyear{2019}

\begin{document}
\label{firstpage}
\pagerange{\pageref{firstpage}--\pageref{lastpage}}
\maketitle

\begin{abstract}
We implement a detailed dust model into the L-Galaxies semi-analytical model which includes: injection of dust by
\tTwo\ and \tOneA\ supernovae (SNe) and AGB stars; grain growth in molecular clouds; and destruction due to
supernova-induced shocks, star formation, and reheating.  Our grain growth model follows the dust content in molecular
clouds and the inter-cloud medium separately, and allows growth only on pre-existing dust grains. At early times, this
can make a significant difference to the dust growth rate.  Above $z\sim8$, \tTwo\ SNe are the primary source of dust,
whereas below $z\sim8$, grain growth in molecular clouds dominates, with the total dust content being dominated by the
latter below $z\sim6$. However, the detailed history of galaxy formation is important for determining the dust content
of any individual galaxy. We introduce a fit to the dust-to-metal (DTM) ratio as a function of metallicity and age,
which can be used to deduce the DTM ratio of galaxies at any redshift. At $z\lesssim3$, we find a fairly flat mean
relation between metallicity and the DTM, and a positive correlation between metallicity and the dust-to-gas (DTG)
ratio, in good agreement with the shape and normalisation of the observed relations. We also match the normalisation of
the observed stellar mass -- dust mass relation over the redshift range of $0-4$, and to the dust mass function at $z=0$. Our results are important in interpreting observations on the dust content of galaxies across cosmic time, particularly so at high
redshift.
\end{abstract}


\input intro
\input model
\input dustmodel
\input growth

\input content
\input conc

\section*{Acknowledgements}

The authors would like to thank Gerg\"{o} Popping and Phil Wiseman for useful 
discussions during the undertaking of this project and Pierre OCVIRK for his helpful 
correspondence while using our fitting function. We would also like to thank the
anonymous referee for a constructive and detailed report that has significantly 
enhanced the strength of this paper. Much of the data analysis 
was undertaken on the {\sc Apollo} cluster at Sussex University. The authors 
contributed in the following way to this paper. APV undertook the vast majority
of the data analysis and produced the figures; APV and PAT worked on developing 
the grain growth framework building on the model provided by BMBH; SJC created the dust yield tables required for stellar 
dust production with the help of RMY, integrated the dust production and destruction 
framework into \lgal\,, did the intial data analysis and produced the first draft of the paper. 
PAT \& RMY helped APV with the interpretation of the results and structuring the paper. 
PAT \& SMW jointly supervised APV \& SJC, and initiated the project. All authors helped to proof-read the text.

APV 
acknowledges the support of of his PhD studentship from UK STFC DISCnet. 
SJC acknowledges the support of his PhD studentship from the STFC.  PAT 
acknowledges support from the Science and Technology Facilities Council (grant
number ST/P000525/1). BMBH 
was supported by Advanced Grant 246797 ``GALFORMOD''
from the European Research Council and by a Zwicky Prize fellowship.

DustPedia is a collaborative focused research project supported by the European Union 
under the Seventh Framework Programme (2007-2013) call (proposal no. 606847). The 
participating institutions are: Cardiff University, UK; National Observatory of Athens, 
Greece; Ghent University, Belgium; Universit\`e Paris Sud, France; National Institute for
Astrophysics, Italy and CEA, France.  We have benefited greatly from the
publicly available programming language {\tt python}, including the {\tt numpy},
{\tt matplotlib}, {\tt scipy} and {\tt h5py} packages.

\bibliographystyle{mnras}
\bibliography{dust} 

%
\appendix

\section*{Supporting Information}
The required code to run the {\sc L-Galaxies} version with the dust model can be found at \url{https://github.com/aswinpvijayan/L-Galaxies_Dust}.\\
Also the code to make the plots seen here can be found at
\url{https://github.com/aswinpvijayan/LGalaxies_paper_plots}. 

\input acc

\bsp	
\label{lastpage}
\end{document}

%% file: intro.tex
\section{Introduction}
Dust has a major impact on the observed properties of galaxies with almost 30\% of all photons in
the Universe having been reprocessed by dust grains at some point in their lifetime
\citep{Bernstein2002}.  These grains can form in the stellar winds around AGB and other evolved
stars, in supernovae remnants (SNR), and can grow \textit{in situ} within molecular clouds. Processes that
destroy or alter dust grains include shock heating by supernovae, photo-evaporation and chemical
explosions \citep{deBoer1987,Savage1996}.  The dust content of a galaxy thus depends in a complex
way upon the evolutionary history of its interstellar medium.

The purpose of this paper is to implement a model for dust growth and destruction within the
\lgal\ semi-analytic model in order to investigate the evolution of the dust content of galaxies,
with particular regard to the high-redshift Universe.

\subsection{Dust production and destruction}
The stellar sources of dust are, in order of importance, \tTwo\ SNR, AGB stars and \tOneA\ SNR.
These dust yields are dependent upon the age and metallicity of the stellar populations.  For SNR we
use the prescription of \citet{Zhukovska2008} and for AGB stars the tables of \citet{Ferrarotti2006}
-- this is described in detail in \Sec{dustYields} below.  We note that at very high redshifts,
$z\gtrsim 6$ observations in the far-infrared have started to identify dust masses substantially in
excess of the amount formed from SNR and AGB stars \citep[e.g.~][]{Mancini2015,daCunha2015}. It is possible, therefore, that the
dust yields may be higher at earlier times, perhaps due to higher survival rates of dust produced in
SNR \citep[e.g.~][]{Dwek2014} -- we do not consider that here.

It is now generally accepted that, at later times, the dominant source of dust in the Universe is
grain growth inside molecular clouds \citep[e.g.~][]{Mattsson2015}.  Our dust growth model,
described in \Sec{dustGrowth}, builds on that of \citet[hereafter
\citetalias{Zhukovska2008}]{Zhukovska2008} and \citet[hereafter
  \citetalias{Popping2017}]{Popping2017}.  Unlike earlier works, we use a variable limit for the
fraction of an element that can be locked up in dust, motivated by the chemistry of the ISM, and we
explicitly follow the dust growth in molecular clouds and the diffuse inter-cloud medium separately,
finding that the two can be quite different in certain regimes.

Dust is destroyed by sputtering at high temperatures.  In our model, we follow the prescription of
\citet{McKee1989} for dust destruction in SNR, described in \Sec{dustDestruction}, and we consider
dust to be instantly destroyed if it is reheated out of the cold ISM to join the hot corona
of the galaxy.  We ignore other processes, such as interaction with cosmic rays, or ejection from
the cold ISM by feedback from an active galactic nucleus -- we will show in \Sec{dustMF} that we have
an excess of dust in massive galaxies at low redshift and this may be one possible cause of that.

\subsection{Previous modelling}

In recent years, detailed chemical enrichment models have been implemented into both semi-analytic
models \citep[SAMs, e.g.][]{Arrigoni2010,Yates2013,DeLucia2014}, and hydrodynamical simulations
\citep[e.g.][]{Wiersma+09b,Vogelsberger+13,Pillepich2018}, and a detailed modelling of the dust chemistry is the
natural next step. Lately there have been works that incorporated dust models in simulations.

The current efforts of modelling dust in semi-analytic models and hydrodynamic simulations builds heavily 
upon the initial `one-zone' models, first implemented in \cite{Dwek1998} and followed up by \cite{Inoue2003}, 
\cite{Morgan2003}, \citetalias{Zhukovska2008}. The most detailed semi-analytic (SA) work, which we use as a basis for our 
own modelling, is that of \citetalias{Popping2017}, which uses the {\sc SantaCruz} \citep{Somerville1999} 
SA model. Their model was run on a grid of haloes for a range of virial masses with trees created using the extended
Press Schechter formalism; whereas our model uses the full set of trees from the relatively
low-resolution but cosmologically representative Millennium \citep{Springel2005}, and the
higher-resolution Millennium II \citep{Boylan-kolchin2009} simulations (hereafter MR and MRII
respectively).  Where appropriate, we will make comparison to \citetalias{Popping2017} in the
results presented below.

Recent studies \citep[\eg][]{Bekki2013,Mancini2016,McKinnon2016,McKinnon2016a,McKinnon2018,Aoyama2017,Aoyama2018,Gjergo2018,Dave2019,Hou2019} have implemented mechanisms for tracking dust production and destruction in
hydrodynamical simulations. \cite{McKinnon2016,McKinnon2016a} implemented a simplified dust model in the moving mesh
code \textit{AREPO} to investigate dust formation in a diverse sample of galaxies, accounting for
thermal sputtering of grains. Their model gives results in rough agreement at low redshifts for the
dust mass function, cosmic dust density and the mean surface density profiles. In
\cite{McKinnon2018}, the model was improved to track the dynamical motion and grain-size evolution of
interstellar dust grains. They predict attenuation curves for galaxies which show large offsets
from the observed ones. \cite{Aoyama2017,Aoyama2018,Hou2019} considered a simplified model of dust grain
size distribution by representing the entire range of grain sizes with large and small grains. They
find the assumption of a fixed dust-to-gas (DTG) ratio to break down for galaxies older than 0.2 Gigayears
(Gyrs) with grain growth through accretion contributing to a non-linear rise.

\subsection{Observational summary}

To compare simulations with observational data, it is important to understand how observers
calculate the dust properties of their galaxy populations. Derivations of physical dust quantities
are generally done using spectral energy distribution (SED) modelling. Many observational studies of
dust mass \citep[e.g.][]{Casey2014,Clemens2013,Vlahakis2005} in galaxies fit single or multiple
greybodies to galaxy SEDs by assuming an emissivity index, $\beta$ and a dust temperature, T$_d$,
which is quite useful when the available data is limited. More complicated models can also take into
account microscopic dust properties, such as the composition and grain size \citep{Zubko2004}. These
models also typically assume that the properties and conditions are uniform throughout the galaxy
\citep{RR2014}, or that the properties in all galaxies at all times are the same as in the local
Universe \citep{Santini2014}.  For all these reasons, it should be appreciated that measurements of
dust mass come with large systematic uncertainties up to a factor of 2-3 (\citealt{Galliano2011,Dale_2012}).

At higher redshifts, far infrared (FIR), millimetre (mm) and sub-millimetre (sub-mm) observations are generally only
possible in extreme galaxies, such as those undergoing starbursts or heavy AGN activity. Sub-mm and mm observations have
been shown to be powerful tools in determining how dust and gas are evolving in high-redshift galaxies, with molecular
transitions such as CO and the continuum emission used to determine the properties of gas
\citep[e.g.][]{Greve2005,Tacconi2006,Scott2011} and dust \citep[e.g.][]{daCunha2008} respectively. Sub-mm observations
are extremely good at tracing the cold dust component of the galaxy which usually dominates the dust mass. ALMA
observations have been instrumental to systematically map the dust continuum
\citep[\eg][]{Hodge2013,Scoville_2016,Dunlop2017,Franco2018} and in some cases, where multi-wavelength data is
available, the dust content of galaxies at redshifts of 2--4 \citep[\eg][]{daCunha2015}. Further complications arise
from the further heating of dust at higher redshifts due to the CMB \citep{daCunha2013}, and the lack of many
observational data points in the FIR means that a dust temperature can not be calculated from the SED and one must be
assumed. The assumption of a dust temperature can lead to differing dust masses by up to an order of magnitude
\citep{Schaerer2015}.

The observational study of local galaxies by \cite{RR14_DTG} found that the dust-to-metal ratio (DTM) is approximately constant in the majority of galaxies. However, at low metallicities, the ratio decreases, suggesting that dust destruction wins out over grain growth.  Also at low redshift, \cite{DeVis2019} found that the DTM ratio of DustPedia galaxies \citep[see][]{Davies_2017} increases as galaxies age, before becoming approximately constant once the gas fraction drops below 60\,\%. For galaxies at higher redshifts ($z>1$), the DTM ratio is seen to increase with metallicity over the broad redshift range of $2 \lesssim z \lesssim 5$ \citep[\eg][]{DeCia2016,Wiseman2017}, again suggesting that a significant amount of dust is formed due to \textit{in situ} grain growth in the ISM.

There also have been detections from deep ALMA and PdBI observations of galaxies at extremely high redshifts ($z>6$) with large reservoirs of dust
($>10^{8}\,\Msun$) \citep[\eg][]{Mortlock2011,Venemans2012,Watson2015,daCunha2015}.  Models to reproduce these \citep[\eg][]{Michalowski2015,Mancini2015} require either enhanced dust
production from supernovae and AGB stars (and reduced destruction by the former), or very rapid dust production soon
after chemical enrichment, suggesting very short grain growth timescales in these metal-poor environments. We will look at all these aspects of the dust evolution paradigm in the following sections.

\subsection{Structure of the paper}

This paper is structured as follows: in \Sec{model} we describe briefly the \lgal\ SA model and some
of the key ingredients that have been incorporated, including the new two phase-model of cold
ISM. In \Sec{dust} we introduce our dust model and describe how it is implemented.  We
present our results on dust growth in \Sec{resultsGrowth}, and of the dust content of galaxies in
\Sec{resultsContent}. Finally, we present our conclusions in \Sec{conc}.

Throughout this paper we adopt the initial mass function (IMF) of \cite{Chabrier2003}, assume the
cosmological parameters derived by {\it Planck} \citep{PlanckCollaboration2014} and use a solar 
metallicity value, $Z_{\odot}$ = 0.0134 \citep{Asplund2009}. 

%% file: model.tex
\section{The Model}
\label{sec:model}

Semi analytic models provide a relatively inexpensive method of self-consistently evolving the
baryonic components associated with dark matter merger trees, derived from $N$-body simulations or
Press-Schechter calculations.  The term semi-analytic comes from the use of coupled differential equations to capture the essential galaxy formation physics determining the properties of gas and stars.  Most modern SAMs include descriptions of: primordial infall and the impact of an ionizing UV
background; radiative cooling of the gas; star formation recipes; metal enrichment; super-massive
black hole growth; supernovae and AGN feedback processes; and the impact of the environment and
mergers on galaxy morphologies and quenching.

\lgal, \citep[and references therein, hereafter \citetalias{Henriques2015}]{Henriques2015} has been
developed over the years to include most of the relevant processes that affect galaxy evolution. In
this work we use a modified version of that model which includes: detailed chemical enrichment (\Sec{modelChem}); the
differentiation of molecular and diffuse atomic phases in the cold gas (see \Sec{modelMol}); and the
detailed dust model introduced in this paper (see \Sec{dust}). We highlight the changes relevant to
our dust model below. An overview of all the physics contained within the \citetalias{Henriques2015}
version of the model can be found in the appendix of that paper.

\noindent The main non-standard symbols used in our model are:
\begin{shortitem}
\item $\mu$ -- fraction of the cold ISM that is in molecular clouds;
\item $f$ -- fraction of metals within molecular clouds which condenses into dust;
\item $g$ -- fraction of metals within the diffuse inter-cloud medium which condenses into dust.
\end{shortitem}

\noindent When describing the dust content, we use the following subscripts:
\begin{shortitem}
\item d -- total amount of dust; 
\item $j$ -- elements;
\item $x$ -- dust species.
\end{shortitem}
\subsection{Detailed chemical enrichment}
\label{sec:modelChem}

Many galaxy formation models use an instantaneous recycling approximation that assumes that stars
pollute their environments with metals the moment they are born. Given the long lifetimes of
low-mass stars, this will introduce too many metals (and thus too much dust) at very early times.
The detailed chemical enrichment model used here \citep{Yates2013} only injects metals into
the environment at the end of a star's life. The model takes the metal production rate from stellar
mass and metallicity dependent yield tables for type-II supernovae \citep{Portinari1998}, type-Ia
supernovae \citep{Thielemann2003}, and AGB stellar winds \citep{Marigo2001}.

As discussed in \citet{Yates2013}, we follow the prescription of \citet{Tinsley1980} for the total
rate of metals $j$ ejected by a stellar population at a time $t$:
\begin{equation}
\label{eq:metal_ejecta_rate}
\dot{M}_j(t) = \int_{M(t)}^{M_{\rm up}} M_j(M,Z_0)\,\psi(t-\tau_{\rm{M}})\,\phi(M) \,\rm{d}M.
\end{equation}
Here $M_j(M,Z_0)$ is the mass of metals released by a star of mass $M$ and initial metallicity
$Z_0$, $\psi(t-\tau_{\rm{M}})$ is the star formation rate at the time of the star's birth, and
$\phi(M)$ represents the normalised initial mass function (IMF) by number. The lower limit of
the integration, $M(t)$, is the mass of a star with a lifetime $t$ (which would be the lowest mass
possible to have died by this time), and the upper limit, $M_{\rm up}$, is the highest mass star
considered in this work, which is 120M$_\odot$.

The stellar lifetimes used in the chemical enrichment calculations are taken from the
\citet{Portinari1998} mass and metallicity-dependent tables. These provide the lifetime of stars of
mass $0.6 \leq \rm{M/M}_\odot \leq 120$ and for five different metallicities ranging from Z = 0.0004
to 0.05.

With this chemical enrichment model incorporated, \textsc{L-Galaxies} is able to simultaneously reproduce a range
of observational data at low redshift, including the mass-metallicity relation for star-forming
galaxies, the abundance distributions in the Milky Way stellar disc, the alpha enhancements in
the stellar populations of early-type galaxies, and the iron content of the hot intra-cluster medium (see \citealt{Yates2013,Yates2017}).

\subsection{Molecular gas}\label{sec:modelMol}

The standard \lgal\ model does not differentiate between atomic and molecular hydrogen in the cold
ISM.  To model this, we implement the molecular hydrogen prescription used in \cite{Fu2013} to split
the cold gas medium into two components - the diffuse ISM and molecular clouds,  
based on the fitting equations in \cite{McKee_2009}. In that model, the
molecular gas fraction $\mu$ is given by
\begin{equation}\label{eq:fh2}
	\mu =
	\begin{cases}
	\dfrac{4-2s}{4+s},& s < 2;\\
	0,& s \geq 2.
	\end{cases}
\end{equation}
The parameter $s$ in Equation \ref{eq:fh2} is defined as
\begin{equation}
	s = \frac{\tn{ln}\big(1+0.6\chi+0.01\chi^2\big)}{0.6\tau_c}
\end{equation} 
where $\chi=0.76(1 + 3.1Z^{'\,0.365})$ and $\tau\sub{c} =
0.066\,(\Sigma\sub{comp}/\tn{M}_{\odot}\tn{pc}^{-2})\,Z^{'}$, with $Z^{'} = Z\sub{gas}/Z_{\odot}$
being the gas-phase metallicity (including metals locked up in dust) relative to the solar value. Also,
\begin{equation}\label{eq:clumping}
	\Sigma_{\rm{comp}}=c_{\rm{f}}\Sigma_{\rm{gas}}
\end{equation} 
where $\Sigma\sub{gas}=M\sub{cold}/\pi r\sub{d}^{2}$ is the surface density, $r\sub{d}$ is the galaxy disk scale length, and $c_{\rm f}$ is a metallicity-dependent clumping factor given by
\begin{equation}
  c_{\rm{f}} =
  \begin{cases}
    0.01^{-0.7},& Z'\leq0.01;\\
    Z^{'-0.7},& 0.01<Z'<1;\\
    1,& Z'\geq 1.
  \end{cases}
\end{equation}
which is meant to account for starburst systems in low-metallicity dwarf galaxies.

In our new model, supernovae and stellar winds are assumed to inject a fraction $(1-\mu)$ of their metal and dust into the
diffuse component and a fraction $\mu$ into the molecular cloud component. However, star formation and dust growth on
grains occurs only in molecular clouds.

We also note that our results remain unchanged on using the molecular hydrogen partitioning recipe used in \cite{Martindale2017} implementing a partitioning based on the mid-plane hydrostatic pressure in the galactic disc from \cite{Blitz2006}.

%% file: dustmodel.tex
\section{Detailed dust model}
\label{sec:dust}

In this section, we describe the new detailed dust model that we have incorporated into \textsc{L-Galaxies}. Our model
traces the three dominant sources of dust production in the Universe; injection by \tOneA\ and \tTwo\ supernovae,
stellar winds from AGB stars, and the growth of dust within molecular clouds. We also implement a model of dust
destruction induced by supernovae shocks and gas heating. We make the assumption that dust grains only reside within the
cold ISM, as the temperature in the hot circumgalactic and intra-cluster media around galaxies is sufficiently high that
dust grains will be rapidly destroyed in those gas phases. This is an oversimplification as dust is observed in both the
CGM \citep[\eg][]{Peek_2015} and ICM \citep[\eg][]{Gutierrez_2014}. \cite{Tsai1995} adopted an analytic form for the decrease in the dust grain radius in the hot phase. The sputtering timescale derived from this (used in other studies, \eg \citealt{McKinnon2016a,Aoyama2018}) can vary between 1 Myr - 10 Gyr depending on the temperature and the density of the hot phase. Since the sputtering timescales of dust in the hot phase is
strongly dependent on the assumed model, we do not consider that here and focus on the dust content of the ISM. This
aspect will be revisited in a future work. 

The dust production rate of a galaxy is therefore
\begin{equation} 
\label{eq:dust_main}
\dot{M}_\dd(t) = \dot{M}_\mathrm{d,inj} + \dot{M}_\mathrm{d,grown} - \dot{M}_\mathrm{d,dest} -
\dot{M}_\mathrm{d,trans},
\end{equation}
where $\dot M_\mathrm{d,inj}$ is the dust yield rate from stellar sources (supernovae and AGB stars), $\dot{M}_\mathrm{d,grown}$ is
rate of dust growth in molecular clouds, $\dot{M}_\mathrm{d,dest}$ is the dust destruction rate, and
$\dot{M}_\mathrm{d,trans}$ is the rate at which dust is transferred out of the cold ISM through
processes such as star formation or mergers.  We discuss each of these processes in more detail
below.

\subsection{Supernova and AGB dust yields}
\label{sec:dustYields}

By analogy to \Eq{metal_ejecta_rate} we have
\begin{equation} 
\label{eq:dust_yield}
\dot{M}_\mathrm{d,inj} = \int_{M(t)}^{M_{\rm up}} M_{\rm{d}}(M,Z_0)\,\psi(t - \tau_M)\,\phi(M)\,\dd M,
\end{equation}
where $M_{\rm{d}}(M,Z_{0})$ is the mass of dust produced by a star of mass M and initial metallicity
$Z_{0}$, and the other parameters are as described in \Sec{modelChem}.  We apply this equation for
both AGB winds from lower-mass stars and for supernovae.

The mass of dust produced by a low mass star of given mass and metallicity (i.e. AGB stars) is taken
from the tables of \cite{Ferrarotti2006}. In this case, the upper limit of the integral is the maximum
possible mass for an AGB star, which is about 8M$_\odot$.

For supernovae, we follow the prescription laid out in \citetalias{Zhukovska2008}.  There it is 
assumed that the mass of dust formed in a supernova remnant is proportional to the total mass 
return of the key element required to form that particular type of dust. The four types of dust 
they consider are silicates, carbon, iron, and silicon carbides, where the key element that 
comprises each species is Si or Mg, C, Fe, and Si, respectively.

We use the following equation to govern the production rate of dust formed by supernovae for the
four separate dust species (denoted by a subscript $x$) that we consider:
\begin{equation}
\label{eq:SNe_yields}
\dot{M}_x = \eta_x \dot{M}_j \frac{A_x}{A_j},
\end{equation}
where $\dot{M}_j$ is the mass return rate of the key element, which we obtain from our detailed
chemical enrichment model as described in \Sec{modelChem}, and $A_x$ and $A_j$ are the atomic
weights of the dust species and key element, respectively. The condensation efficiency
parameter, $\eta_x$, is used for converting a specific element into dust, as estimated from observations of local
supernovae remnants. These efficiency parameters are defined considering the effects of the reverse
shock and are therefore smaller than they would be for initial dust condensation. 

We apply \Eq{SNe_yields} to all four different dust species for \tTwo\ supernovae, and for
iron-based dust from \tOneA\ supernovae. The values of the parameters that we use are given in
\Tab{sne_eff}.
\begin{table}
	\centering
	\caption{The conversion efficiencies used for the production of dust grains in supernovae remnants based on the mass return of key metals. The efficiencies have been adopted from \citetalias{Zhukovska2008}.}
	\label{tab:sne_eff}
	\begin{tabular}{l|llll}
		\hline
		& \multicolumn{4}{c}{Dust Species ($x$)} \\
		& silicates & carbon & iron  & SiC    \\ \hline
		$\eta_{\rm{x,SNII}}$        & 0.00035     & 0.15   & 0.001 & 0.0003 \\
		$\eta_{\rm{x,SNIA}}$        & 0.0       & 0.0    & 0.005 & 0.0    \\
		$A_x$ & 172.0     & 12.01  & 55.85 & 40.10\\ 
		\hline
		& \multicolumn{4}{c}{Key Element ($j$)} \\
                & Si / Mg        & C      & Fe    & Si     \\
                $A_j$	&121.4 / 24.31	&12.01	&55.85	&28.09\\
                \hline
	\end{tabular}
\end{table}

\subsection{Grain growth in molecular clouds}
\label{sec:dustGrowth}

\begin{figure*}
  \begin{center}
    \includegraphics[width=0.95\textwidth]{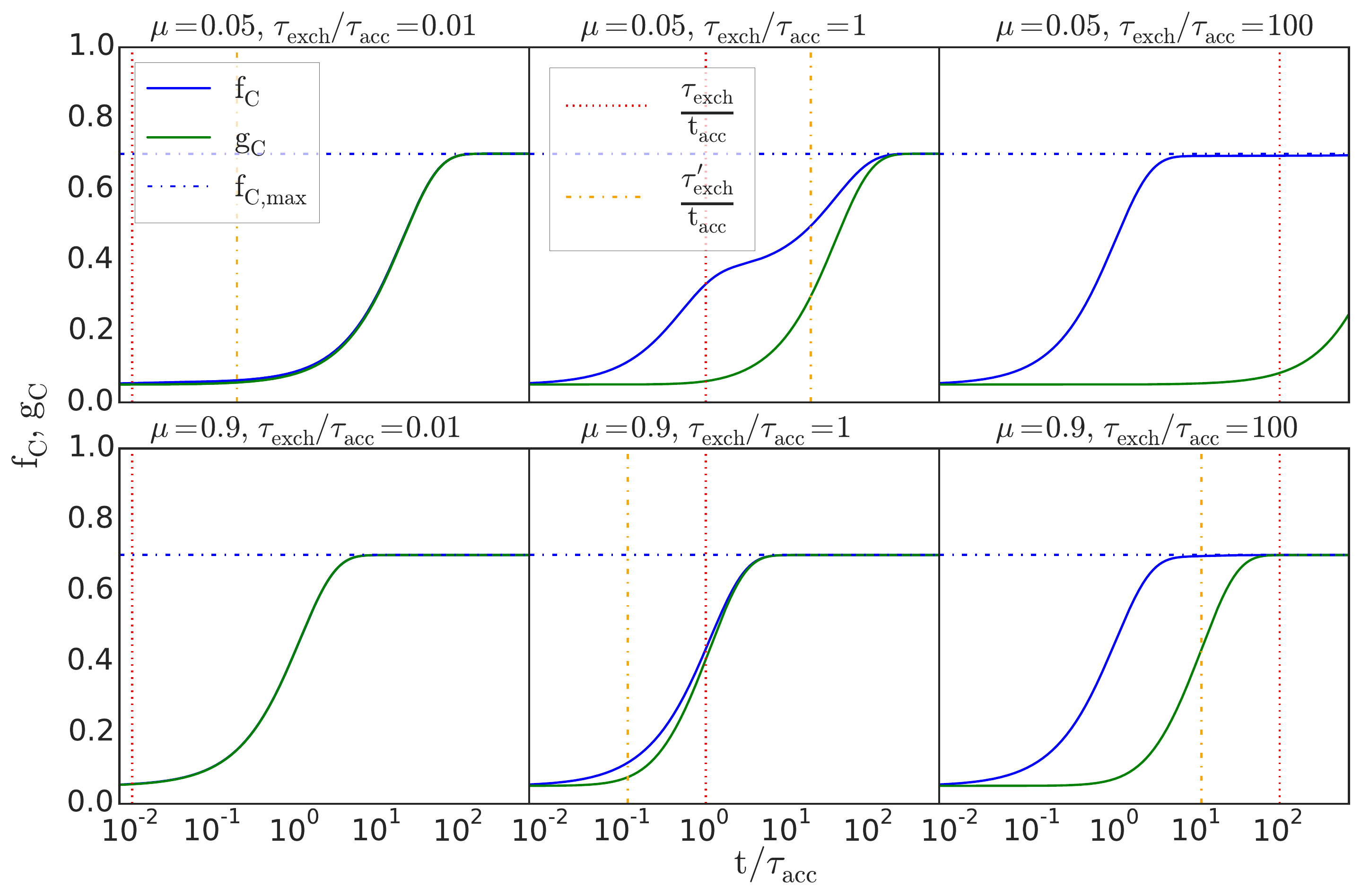}
    \centering
    \caption{Evolution of the fractions $f_\mathrm{C}$ and $g_\mathrm{C}$ for different values of
      $\tau_\mathrm{exch}$ and $\mu$ with an initial dust fraction $f_\mathrm{C} = g_\mathrm{C} =
      0.05$. These plots are valid for a constant value of $\tau_{\rm{acc}}$, which in our model
      decreases with the production of more dust, speeding up the saturation of the two
      fractions. The horizontal dot-dashed line represents the maximum permissible condensation
      value, fixed here at 0.7 for carbon.  The vertical lines show the ratios of the accretion and
      exchange timescales. }
    \label{fig:fc_fd}
  \end{center}  
\end{figure*}

\begin{figure}
	\begin{center}
		\includegraphics[width=0.47\textwidth]{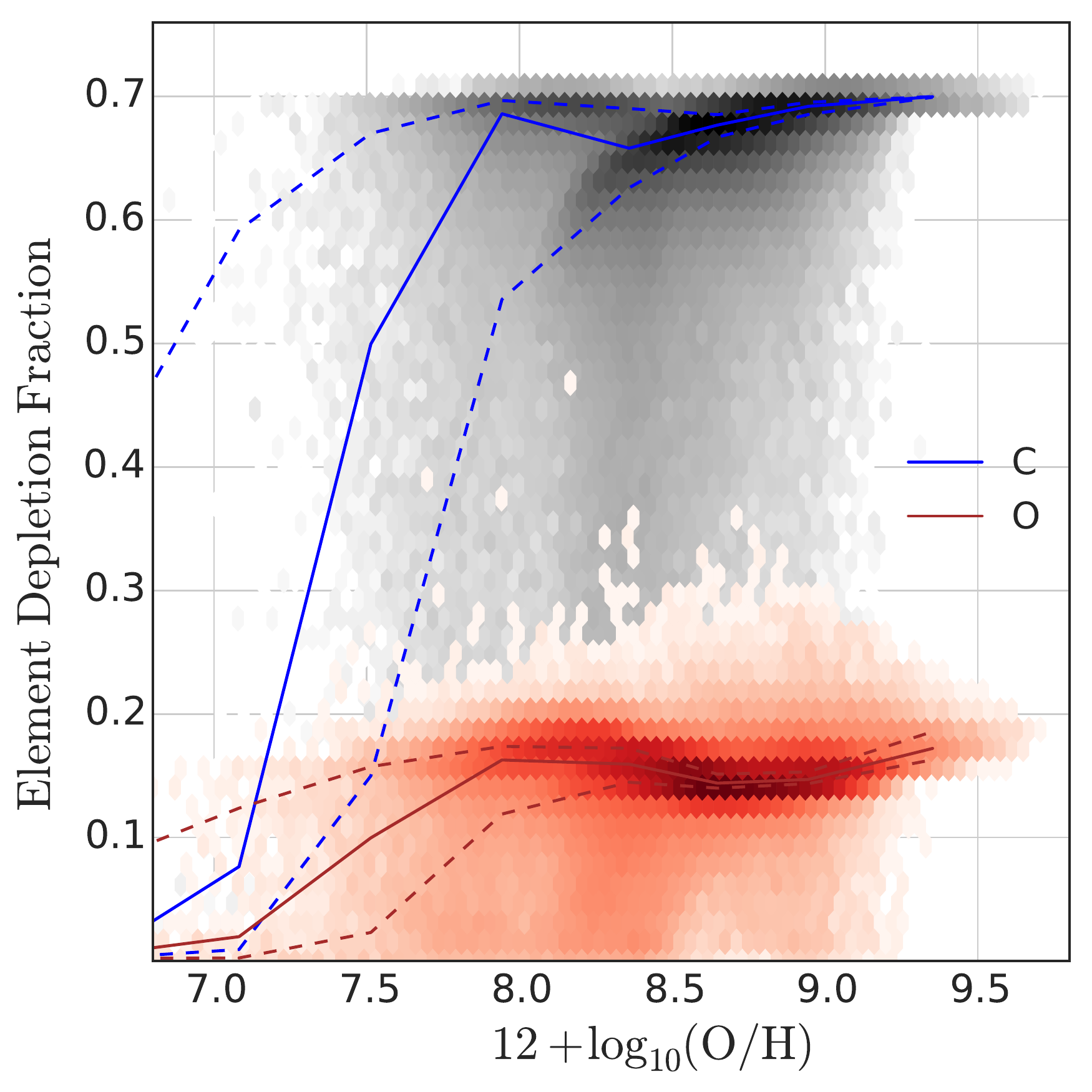}
		\centering
		\caption{Carbon and oxygen depletion fractions plotted against the total cold-gas metallicity for $z=0$. Blue and brown lines denote the median result from galaxies in our model, with the dashed lines denoting the 84 and 16 percentiles.}
		\label{fig:C O depletions}
	\end{center}  
\end{figure}

A complete model for grain growth would consider how the accretion of different elements varies with
different grain sizes, shapes, compositions and grain chemistry, but this would become very
complicated.  Here, we follow \citetalias{Popping2017} in adopting a simpler model in which grain
growth inside molecular clouds occurs on a timescale referred to as the accretion timescale
($\tau_{\rm{acc}}$), and exchange of materials between the molecular clouds and the diffuse media is
governed by an exchange timescale ($\tau_{\rm{exch}}$) which is also the average lifetime of
molecular clouds and is set to 10\,Myr \citep{Zhukovska2014}. 

For each element $j$ in the molecular-cloud component of the ISM, we set a maximum condensation fraction that can be locked up in dust,
$f_{j,\mathrm{max}}$. There is also an implicit maximum $g_{j}$, which is set by the $f_{j,\mathrm{max}}$ in the molecular clouds. This we fix at unity for the refractory elements Mg, Si, Ca, and Fe, while
for N, Ne and S it is set to 0. Neon is unreactive, nitrogen is mostly bound up in volatile
gases and sulphur shows little or no incorporation into dust grains \citep{Jones2000}. In the case of carbon and oxygen, we follow \citetalias{Zhukovska2008} to estimate
$f_{j,\mathrm{max}}$. 
Some carbon is locked up as CO in the molecular clouds and thus not available
for grain growth. Observations estimate the fraction of carbon that is locked up as CO inside
molecular clouds to be around 20-40\,\% \citep{Irvine1987,Dishoeck1998}. In our model we fix this at
30\,\%, thus setting $f_\mathrm{C;max} = 0.7$. In the case of oxygen, we assume it is present in
dust in the form of metal oxides. Thus, the maximum fraction of available oxygen is set by the
amount of other elements present to form these compounds, which are silicates and iron oxides in our
model. Following \citetalias{Zhukovska2008}, we adopt olivine ([Mg$_y$Fe$_{1-y}$]$_2$SiO$_4$) and
pyroxene (Mg$_y$Fe$_{1-y}$SiO$_3$) as the major silicate compounds in the ISM in the ratio 32:68;
here we take $y = 0.8$. In the case of iron oxides, we assume hematite (Fe$_2$O$_3$) and magnetite
(Fe$_3$O$_4$) are the major compounds, contributing equally towards dust growth. Thus for oxygen the
maximum condensed fraction in molecular clouds depends on the chemical composition. 

Grain growth is then implemented by solving the following pair of coupled differential equations at
each timestep within the simulation:
\begin{gather}\label{eq:f}
  \frac{\dd f_j}{\dd t} = \frac{f_{j,\mathrm{max}} - f_j}{\tau_{\rm{acc}}} + \frac{g_j-f_j}{\tau_{\rm{exch}}};
\end{gather}
\begin{gather}\label{eq:g}	
  \frac{\dd g_j}{\dd t} = \frac{f_j-g_j}{\tau_{\rm{exch}}} \frac{\mu}{1-\mu} = \frac{f_j-g_j}{\tau_{\rm{exch}}^\prime}
\end{gather}
where $f_j$ and $g_j$ are the condensation fractions of element $j$ in dust within the molecular
clouds and the diffuse medium, respectively, and $\tau_{\rm{exch}}^\prime$ is the effective exchange timescale over which all the ISM in a galaxy is cycled through molecular clouds (see \citealt{Zhukovska2014}).

Figure \ref{fig:fc_fd} shows how the condensation fractions $f_j$ and $g_j$ evolve for the particular case of
carbon. Columns show different $\tau_{\rm{exch}}/\tau_{\rm{acc}}$ ratios, and rows show different molecular gas fractions, $\mu$. For values of $\tau_{\rm{exch}} \ll \tau_{\rm{acc}}$, the condensation fractions evolve similarly for both high and low $\mu$.  For $\tau_{\rm{exch}} \gg \tau_{\rm{acc}}$, $f_{\rm C}$ saturates at $f_{\rm{C,max}}$ relatively quickly. However, $g_{\rm C}$ takes a much longer time to reach its maximum allowed value, with its evolution being particularly slow in regions with low $\mu$ (\ie{}dominated by diffuse gas).

Because dust catalyses the formation of other dust, we use the following expression for the
accretion timescale, which differs from some of the expressions used in previous studies in that it
uses the dust mass instead of the metal mass in the denominator:
\begin{equation}\label{eq:tacc}
\tau_{\rm{acc}} = \tau_{\rm{acc,0}}\times\bigg(\frac{\textrm{Total mass in clouds}}{\textrm{Mass of dust
    in clouds}}\bigg)
\end{equation}
We require a short cooling time, $\tau_{\rm{acc,0}} \lesssim 5\times$10$^4$ yr to match the high dust masses observed
at high redshift, and we adopt this as our canonical value (note that this is lower than the 15\,Myr used in
\citetalias{Popping2017} because of our use of dust fraction rather than metallicity in the growth equation). The impact
of varying the value of $\tau_{\rm{acc,0}}$, as well as the evolution of $\tau_{\rm{acc}}$ with redshift,
is discussed in Appendix \ref{sec: acc_scale}.

We also show in Figure~\ref{fig:C O depletions} the depletion fraction \ie $M_{j,\mathrm{dust}}/(M_{j,\mathrm{cold}}+M_{j,\mathrm{dust}})$ against total gas-phase ISM metallicity in our model for the case of carbon and oxygen. We find that the typical carbon depletion fraction increases over cosmic time, whereas the typical oxygen depletion fraction maintains a value of 0.1 - 0.2 below $z \sim 4$. Our values are comparable to those adopted by emission-line modelling studies \citep[\eg][]{Groves_2004,Gutkin2016}, and our model reproduces the expectation that oxygen has a relatively low depletion onto dust grains \citep[\eg][]{Jones2000,Jenkins_2009}. 
\subsection{Dust destruction}
\label{sec:dustDestruction}

We implement a model of dust destruction due to the effects of supernovae induced shock waves
following the prescription of \cite{McKee1989}:
\begin{equation}
\label{eq:dustDest}
\dot M_{d,{\rm dest}} = \frac{M_\mathrm{d}}{\tau_{\rm dest}},
\end{equation}
where $\tau_{\rm dest}$ is the timescale for destruction of dust.
\begin{equation} \label{eq:t_dest}
\tau_{\rm dest} = \frac{M_{\rm cold}}{M_{\rm cleared} f_{\rm SN} R_{\rm SN}}
\end{equation}
where $M_{\rm cold}$ is the mass of the cold ISM in the galaxy, and $R_{\rm SN}$ the rate of supernovae \tTwo\ and \tOneA\
going off in the stellar disk, which we directly model. The other two quantities are parameters of the model: $M_{\rm cleared}$ is the amount of
cold gas that is totally cleared of dust by an average supernovae which we fix at a lower estimate from \cite{Hu2019} of
1200 M$_\odot$; $f_{\rm SN}$ accounts for the effects of correlated supernovae and supernovae out of the plane of the
galaxy, and is set to 0.36 \citep{McKee1989,Zhukovska2013,Lakicevic2015}. For a galaxy of similar stellar and cold-gas mass to the Milky Way, this formalism returns a $\tau_{\rm dest}$ in good agreement with the estimates obtained by \citet{Hu2019} for their hydrodynamical simulations of the multiphase ISM in the solar neighbourhood.  

We assume that the destruction mechanisms act equally on all types and locations of dust, so that \Eq{dustDest} can be
applied equally to all dust species. We do not consider dust destruction due to UV radiation, cosmic rays or grain-grain
collisions.

\subsection{Dust transfer} 
\label{sec:dustOther}

In this section, we briefly describe the other physical processes within \textsc{L-Galaxies} that act on
material within the cold gas phase and thus impact the dust content of galaxies.

\subsubsection{Star formation}
Stars form from the material present in their birth clouds. We therefore transfer the dust within molecular clouds to the stellar component in proportion to the total mass of stars formed:
\begin{equation}
\dot{M}_d = - \frac{M_\dd}{M_\mathrm{cloud}} \dot{M}_*
\end{equation}
where $M_\dd$ is the mass of dust within, and $M_\mathrm{cloud}$ the total mass of, the molecular
clouds, and $\dot{M}_*$ is the star formation rate. It should be noted that the star formation prescription is the same as in \cite{Henriques2015}. 

\subsubsection{Mergers}
\lgal\ has separate prescriptions for minor and major mergers. In a major merger, the gas discs of the two progenitor galaxies are assumed to be completely removed through merger-induced star formation and the associated galactic winds driven by supernovae. As we made the assumption that dust can only exist within the ISM, this effectively destroys the dust.

In a minor merger, the disc of the larger galaxy survives and the cold gas component of the smaller
galaxy is accreted onto it. In this case, we assume the dust components of the two merging galaxies
survive and are placed into the respective disc component of the more massive galaxy.

\subsubsection{Other dust destruction mechanisms}
There are several other mechanisms, such as reheating or cooling, that transfer dust between different gas phases 
in a galaxy, such as when supernovae heat up cold gas. Whenever any dust is transferred out of the ISM 
within our model, we destroy that dust and return it to its metal components. Since we assume dust is completely destroyed in the hot phase, no dust gets transferred from hot to cold phase -- this will not significantly alter our results, as there is already a strong equilibrium between the rate of dust production and destruction in the ISM in our current formalism. We direct the reader to 
the appendix of \citetalias{Henriques2015} for a complete description of all the processes that affect the gas
phases.

%% file: growth.tex
\section{Results: Dust Growth}
\label{sec:resultsGrowth}

In this section, we begin to present some of the results of our model regarding the nature and efficiency of dust
growth; in the next section, we will look at the resultant dust content of galaxies.  We run the model using the dark
matter subhalo trees from the Millennium (hereafter MR, \citealt{Springel}) and Millennium-II (hereafter MRII,
\citealt{Boylan-kolchin2009}) N-body simulations of hierarchical structure formation, in order to test our model on a
cosmological volume of galaxies, applying a stellar mass selection cut in the respective simulations. Galaxies
below/above a stellar mass of 10$^9\Msun$ are selected from from MRII/MR, respectively.\footnote{The
  precise choice is unimportant as the two agree over approximately a decade in the stellar mass function.} The disjoint
median lines and hex density on the plots that follow can be attributed to the different volumes of the two
simulations. The analysis is restricted to central galaxies (the most massive galaxy inside the halo virial radius),
unless stated otherwise.

\subsection{Dust-to-Metal (DTM) ratio}
\label{sec:resultsDTM}

The most fundamental diagnostic and test of our model is the dust-to-metal (DTM, M$_{\rm{dust}}$/(M$_{\rm{metals}}+$M$_{\rm{dust}}$)) ratio which
measures the efficiency with which metals are converted in to dust. 

\subsubsection{DTM versus stellar mass}

\begin{figure*}
  \begin{center}
    \includegraphics[width=\textwidth]{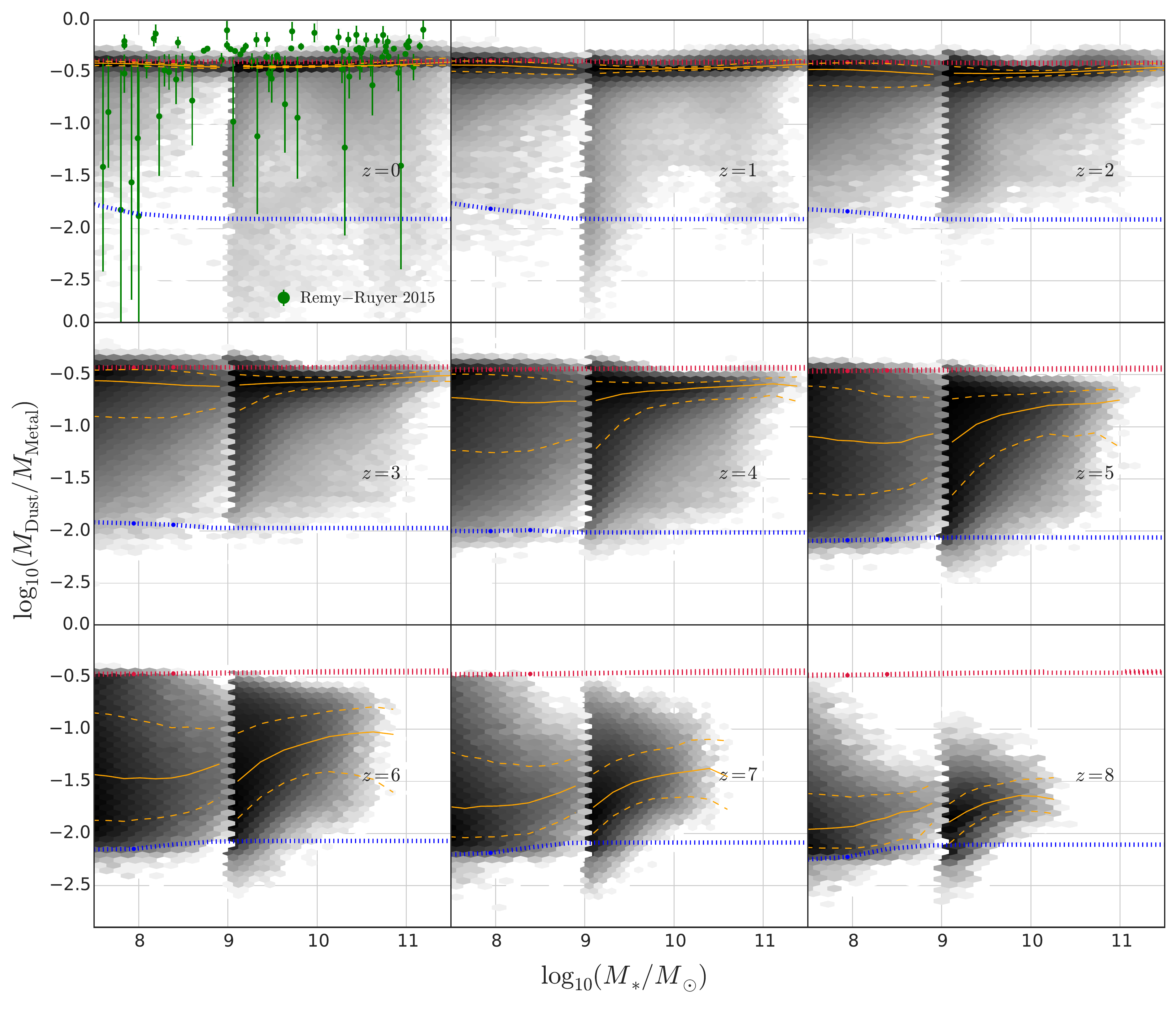}
    \centering
    \caption{The dust-to-metal ratio as a function of stellar mass from $z = 0-8$. The orange line shows the median
      result from galaxies in our model, with the dashed lines denoting the 84 and 16 percentiles. The red dotted line
      represents the saturation limit calculated from average metal abundances in the model while the blue dotted line
      is the median DTM ratio obtained from stellar injection alone. Green points show the observational constraints from \protect\cite{RR2014}.}
    \label{fig:DTM_stellar}
  \end{center}
\end{figure*}

\begin{figure}
	\begin{center}
		\includegraphics[width=0.5\textwidth]{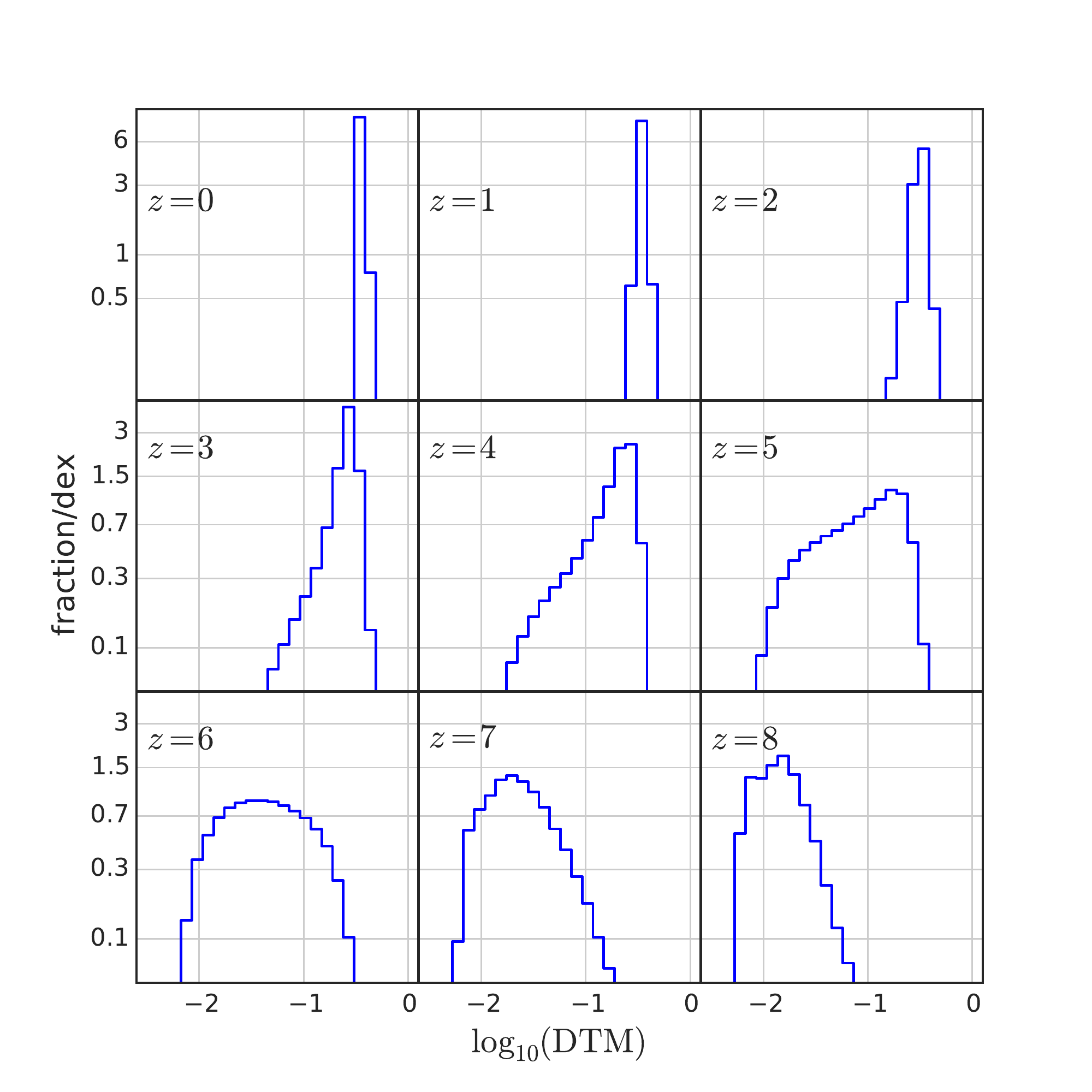}
		\caption{The distribution of DTM ratios for various redshifts from 8 to 0. The peak of the distribution clearly shifts from low to high values over cosmic time.}
		\label{fig:DTM_fractions}
	\end{center}
\end{figure}

\begin{figure*}
  \begin{center}
    \includegraphics[width=\textwidth]{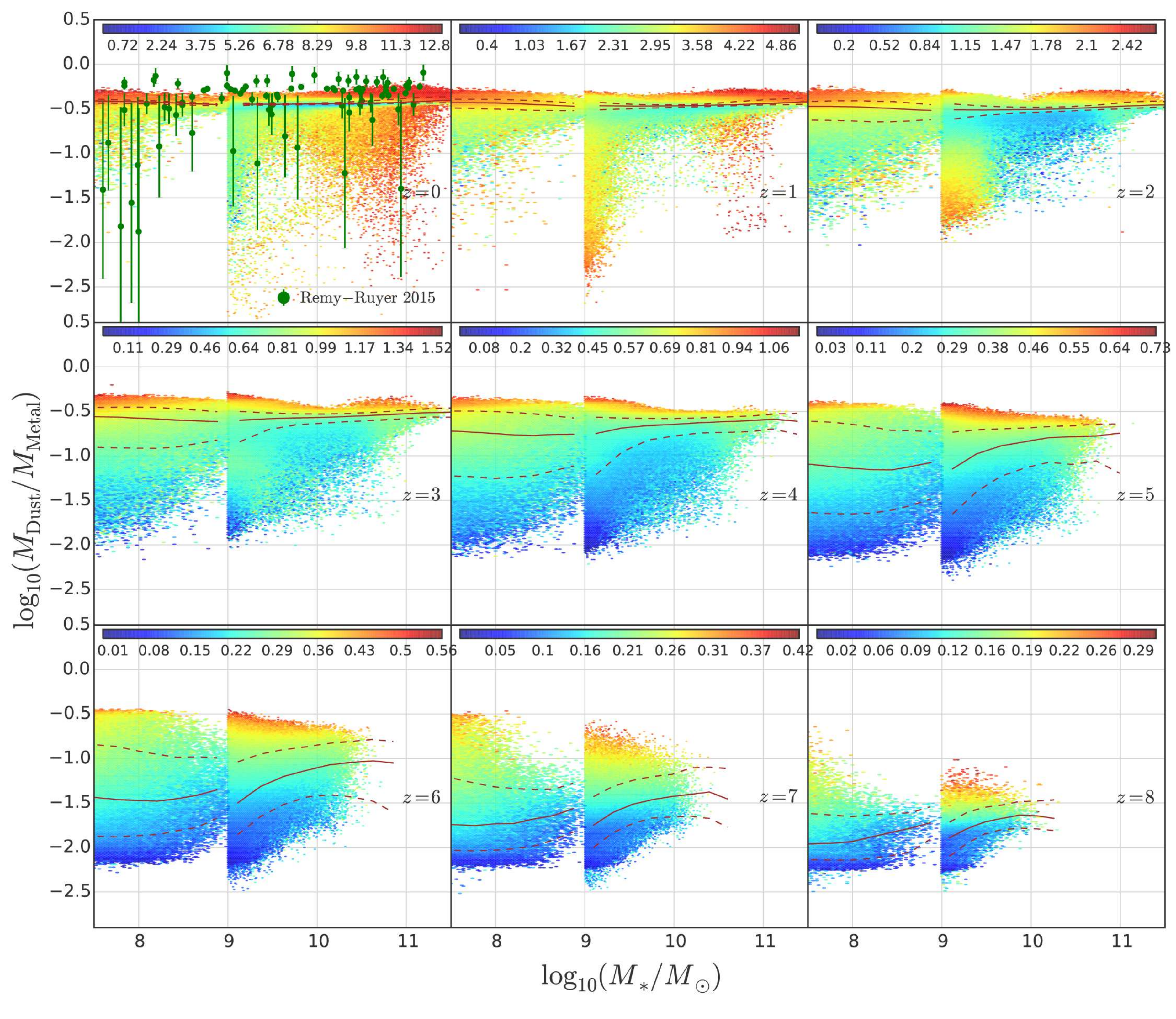}
    \centering
    \caption{The dust-to-metal ratio as a function of stellar mass from $z = 0-8$, same as
      Figure~\ref{fig:DTM_stellar}, with galaxies coloured according to their mass-weighted
      stellar age in Gyrs. Green points show the observational constraints from \protect\cite{RR2014}.}
    \label{fig:DTM_stellar_age}
  \end{center}
\end{figure*}

Figure~\ref{fig:DTM_stellar} shows how the DTM ratio varies with stellar mass, with the green coloured observational
data points taken from \cite{RR2014}. The solid line
shows the median result of the galaxies in our model, while the dashed lines show the 84 and 16
percentiles. The hex diagram in grey shows a 2D density distribution of galaxies in our model. The dotted, red line in the plot shows the maximum possible DTM ratio in our
model (for the median metallicity), assuming that grain growth has saturated (\ie{}$f_{j} = f\sub{\textit{j},max}$ for every element). The
blue dotted line shows the median DTM ratio obtained from stellar dust production mechanisms alone.  The slight
displacement of the median DTM ratio below the saturation value at low redshift is due to dust destruction mechanisms
that offset some of the grain growth; the slight offset of the median DTM ratio above the blue line at high redshift is
due to the fact that dust growth takes off very quickly.
The transition from galaxies dominated by dust injected by stellar sources (mostly \tTwo\ SNe) and that dominated by
grain growth occurs at $z\sim6$, as illustrated in Figure~\ref{fig:DTM_fractions} which shows the fraction of galaxies
in different DTM ratio bins.

The \cite{RR2014} data shown in Figure~\ref{fig:DTM_stellar} combines two samples of local galaxies from the Herschel:
Dwarf Galaxy Survey (DGS \citealt{Madden2013}, to study low-metallicity systems) and the Key Insights on Nearby Galaxies: a Far-Infrared Survey with
Herschel (KINGFISH \citealt{Kennicutt2011}, mostly spiral galaxies along with several early-type and dwarf galaxies to include metal-rich galaxies). They use a semi-empirical dust SED model presented in \cite{Galliano2011}
to derive dust masses and estimate systematic errors of order 2$-$3. The DTM ratio predictions from our model show
reasonable agreement with this data, although the dispersion in the model predictions is lower, and some of the highest
observed DTM ratios are incompatible with the predictions of our model: the extent to which that is due to
observational uncertainty is hard to assess. 

The transition from the lower to the upper locus in Figures~\ref{fig:DTM_stellar} \&~\ref{fig:DTM_fractions} is largely a function of the age of the
galaxy -- grain growth needs time to act (see also Appendix~\ref{sec: acc_scale}).  This is shown clearly in
Figure~\ref{fig:DTM_stellar_age} which plots the same relation with galaxies coloured by their mass-weighted stellar
age. Although the precise time taken for grain growth to saturate will depend upon the metallicity and initial dust
content of the ISM, it takes of order 1\,Gyr to do so. A study by \cite{Inoue2003} has also shown that the evolutionary
tracks in the metallicity -- DTM ratio plane depends on the star-formation history.

At $z = 0$, the DTM ratio in some of the oldest, most massive galaxies has again begun to fall slightly and in some
significantly -- these are early types for which the molecular gas content of the cold ISM is low.
We can therefore see that a galaxy's DTM ratio depends strongly on it's age, but also more weakly on it's chemical enrichment, molecular gas consumption, and other factors relating to its evolutionary history.

If we compare our results to \citetalias{Popping2017} (their Figure 6), our model galaxies do not exhibit any evolution of the DTM ratio with stellar mass as seen in their results at $z=0$. But the scatter at $z=0$, is negligible similar to \citetalias{Popping2017}. At all redshifts their DTM ratio remains almost constant as well as exhibiting negligible scatter below M$_* = 10^{8.5}$M$_{\odot}$, while increasing rapidly afterwards due to their grain growth mechanism dominating the dust production. The cause of these differences are explained in Section \ref{sec:resultsRates}.

\subsubsection{DTM versus metallicity}\label{sec:resultsDTMmet}

\begin{figure*}
  \begin{center}
    \includegraphics[width=\textwidth]{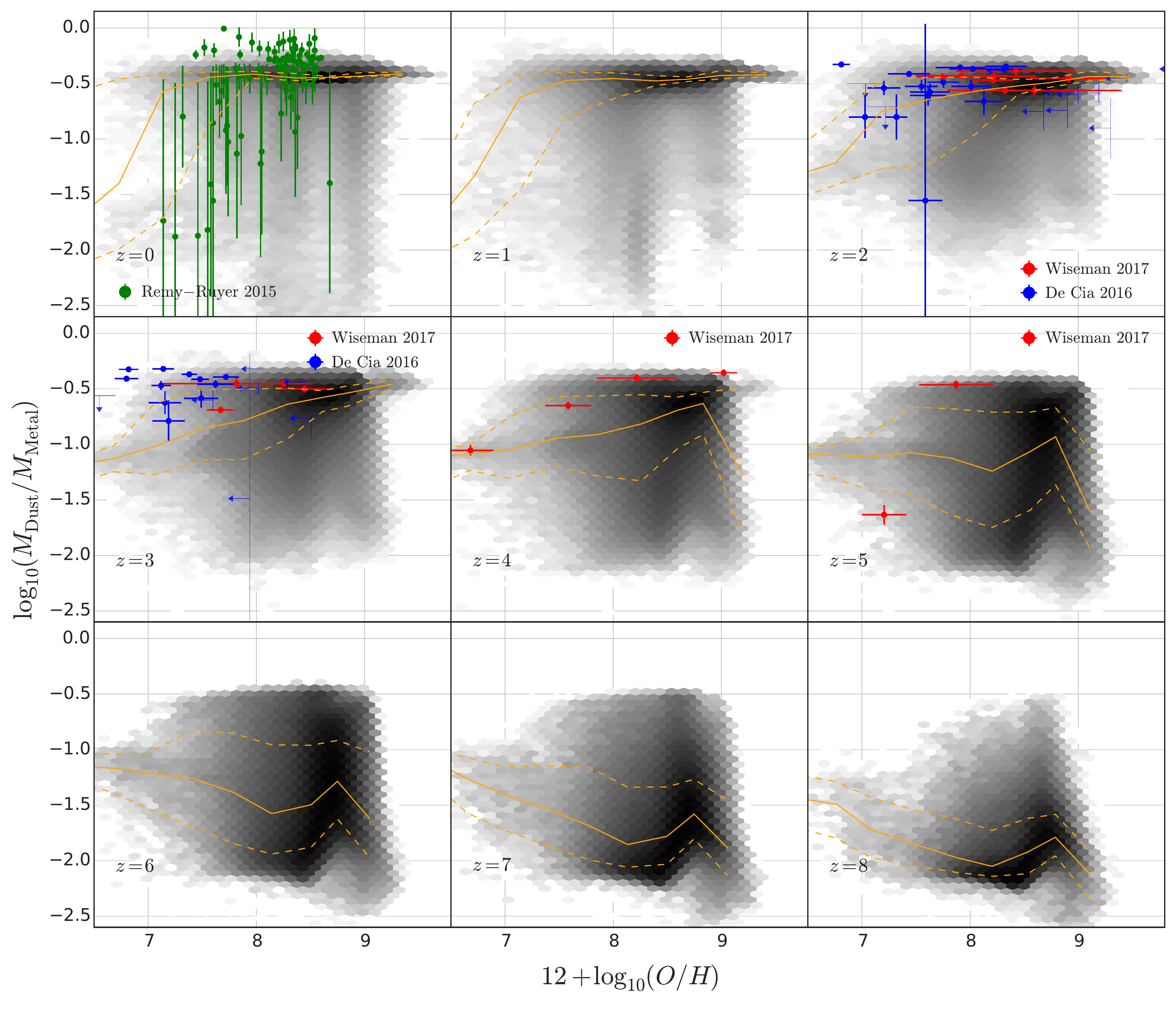}
    \centering
    \caption{The dust-to-metal ratio as a function of their metallicity from z = 0-8. The orange
      line shows the median result from galaxies in our model, with the dashed lines denoting the 84
      and 16 percentiles. Green blue and red points show the observational constraints from
      \protect\cite{RR2014}, \protect\cite{DeCia2016} and \protect\cite{Wiseman2017} respectively.}
    \label{fig:DTM_oxygen}
  \end{center}
\end{figure*}

Figure~\ref{fig:DTM_oxygen} shows the DTM ratio as a function of the gas-phase ISM oxygen abundance (\ie{}the oxygen not
locked up in dust).  At $z = 0$, we again compare to observations from \cite{RR2014}.  We match the normalization of the
observations for $12+\log_{10}$(O/H)$>8$ and also some scatter down to low DTM ratios, noting that the low-DTM observational data tend to have the largest uncertainties. At higher redshifts, we show a good fit to the DTM ratios deduced by observations of gamma ray bursts \citep[GRBs][]{Wiseman2017} and damped lyman-alpha emitters
\citep[DLAs][]{DeCia2016}.

At $z\ge6$ there appears a negative trend in the DTM-metallicity relation with increasing metallicity. This 
is due to the fact that at these high redshifts grain growth has not had sufficient time to enrich the cold ISM. This 
trend also emerges from the dust injection tables used in the model, since at these redshifts the DTM ratio 
follows the stellar injection modes of dust production. The same feature is seen in the model varaints that 
are discussed in \citetalias{Popping2017}. The feature is absent in the fiducial model used in \citetalias{Popping2017} 
due to their grain growth mechanism dominating the dust production (see Section \ref{sec:resultsRates}).

\subsubsection{DTM fitting function}\label{sec:resultsDTMfit}

\begin{figure}
  \begin{center}
  	\centering
    \includegraphics[width=0.48\textwidth]{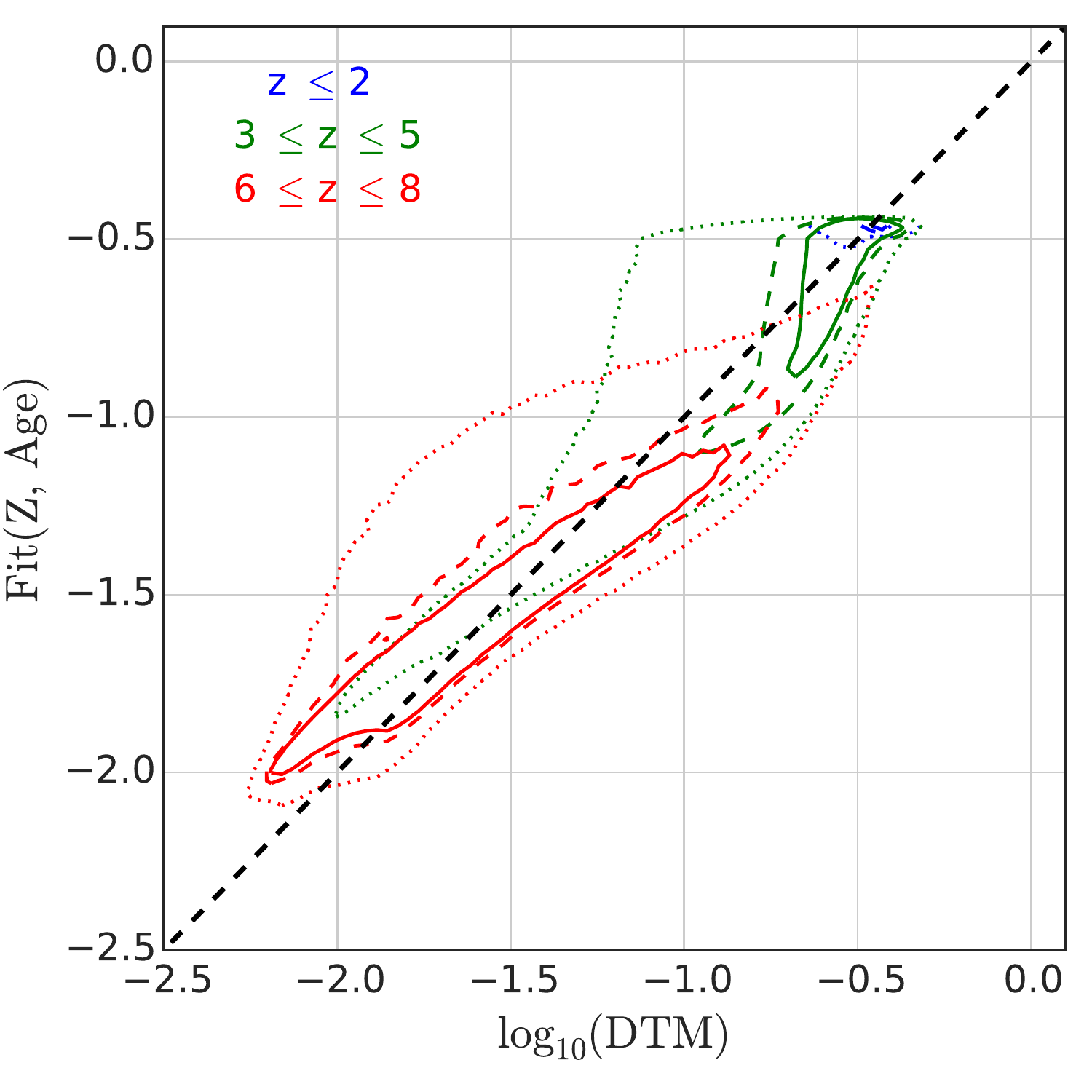}
    \caption{The DTM fitting function in Equation~\ref{eq:DTMfit} is plotted against the DTM ratio
      from the model as contour enclosing different percentiles of the data. The blue, green and red contour lines represent model galaxies with $z\le 2$, $3\le z \le 5$ and $6\le z \le 8$ respectively. Solid, dashed and dotted lines represent the 50, 68 and 95th percentile respectively. The dashed black line represents the one-to-one relation between the fitting 
      function and the data. }
    \label{fig:DTM_fit}
  \end{center}
\end{figure}

As we have seen, the DTM ratio can vary by a large amount, depending upon the evolutionary history of
a galaxy. It would be useful to be able to capture that behaviour with a suitable fitting function. Motivated by our conclusions earlier in this section, we posit the following functional form:
\begin{equation}\label{eq:DTMfit}
\rm{DTM\sub{fit}} = \mathcal{D}_0 + (\mathcal{D}_1-\mathcal{D}_0)\,\big[1-exp\big(-\alpha\,Z^{\beta} (Age/\tau)^{\gamma}\big)\big],
\end{equation}
where $\mathcal{D}_0$ and $\mathcal{D}_1$ represent the initial \tTwo\ SNe dust injection and the saturation value,
respectively, $Z$ is the metallicity of the interstellar medium, Age is the mass-weighted age of the stellar population, and $\tau=\tau_\mathrm{acc,0}/\mathcal{D}_0\,Z$ is an estimate of the initial dust growth timescale after dust injection
from \tTwo\ supernovae but prior to the initiation of dust growth on grains.

Fixing the values of $\mathcal{D}_0$ and $\mathcal{D}_1$ by reference to \Fig{DTM_stellar}, the best fit values (using the Levenberg-Marquardt method implemented in the {\tt python} package {\tt scipy.optimize.curve\_fit}) to the
other parameters are:
\begin{align*}
\mathcal{D}_0 &= 0.008,\\
\mathcal{D}_1 &= 0.329,\\
\alpha &= 0.017,\\
\beta &= -1.337,\\
\gamma &= 2.122.
\end{align*}
The above fitting function is plotted against the DTM ratio in the model in Figure~\ref{fig:DTM_fit} for $z = 0-8$. The
majority of galaxies lie close to the fit, well within about a factor of 2, although the full dispersion in DTM ratios
is not quite captured. This then provides a good estimate of dust extinction should the metallicity and age of a galaxy
be known, and offers a significant improvement upon the fixed DTM ratios often assumed in the literature
\cite[\eg][]{Wilkins2018}.  We show in the appendix that the same fitting function holds good for different choices of
$\tau_\mathrm{acc,0}$.

\subsection{Integrated dust production rates}
\label{sec:resultsRates}

\begin{figure}
  \begin{center}
    \includegraphics[width=0.48\textwidth]{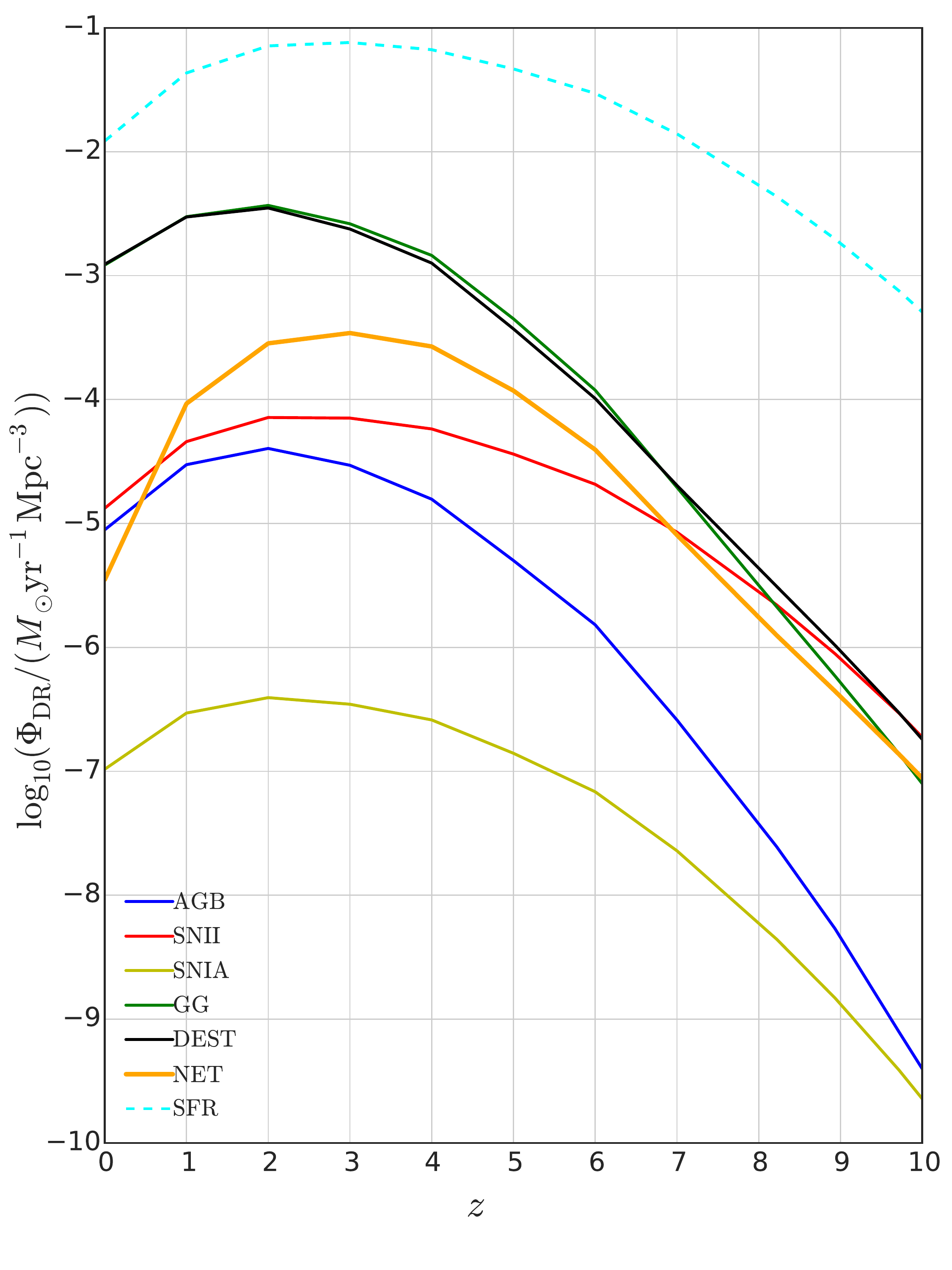}
    \centering
    \caption{The production rate of dust through different mechanisms as a function of redshift for
      the MR run. Red, blue and yellow lines show the contribution from stellar sources of dust
      production, type II supernovae (SNII), AGB stars and type Ia supernovae (SNIA) respectively.
      The green line (largely obscured) shows the contribution from grain growth (GG) inside
      molecular clouds. The black line shows the dust destruction rate (DEST). The orange line shows
      the net total dust production rate (NET), taking into account all production and
      destruction. We also plot the star formation rate (SFR) density as a dashed cyan line for
      comparison. }
    \label{fig:rate_redshift}
  \end{center}
\end{figure}

\begin{figure}
  \begin{center}
    \includegraphics[width=0.48\textwidth]{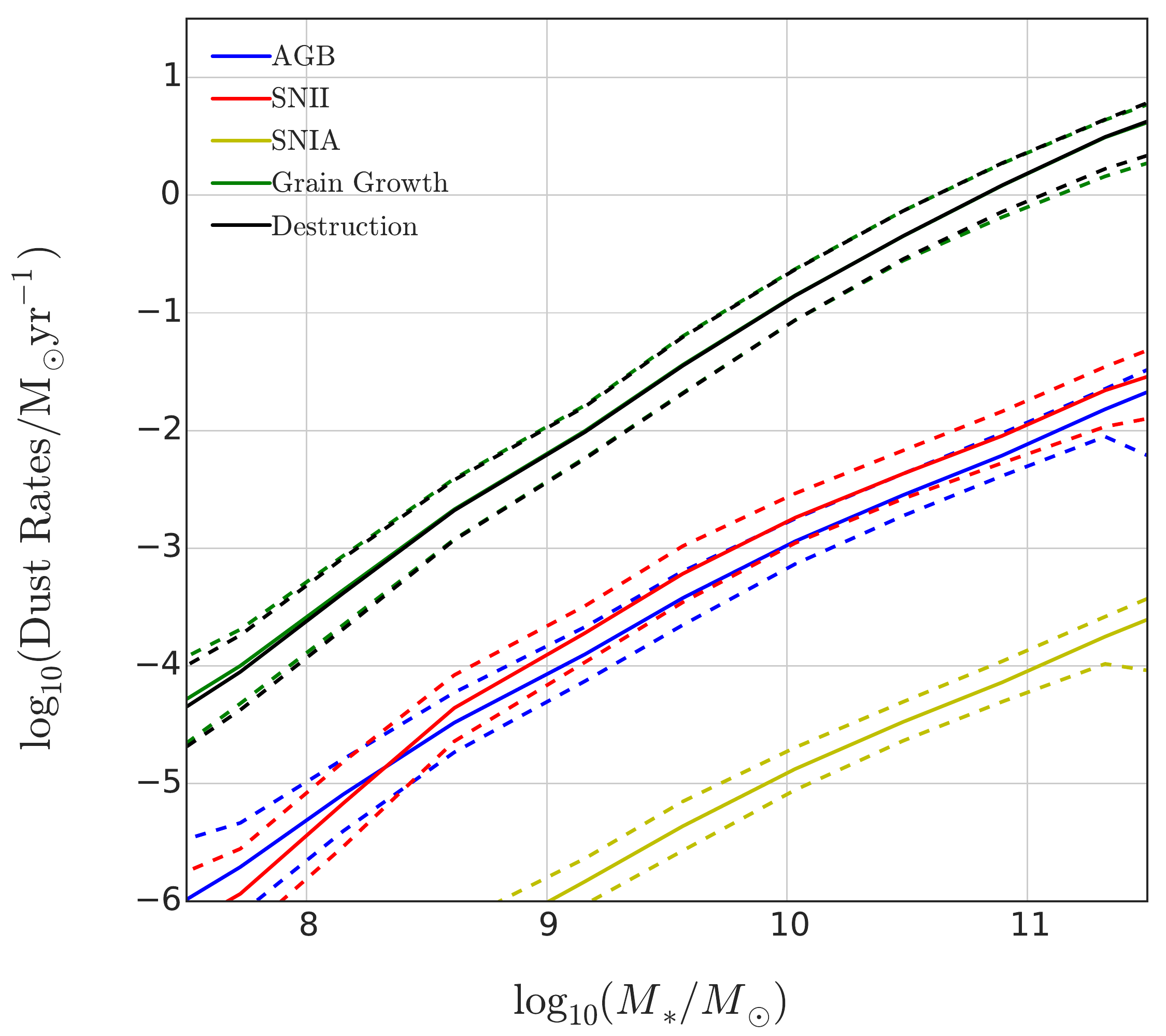}
    \centering
    \caption{The production rate of different dust mechanisms as a function of stellar mass, shown for $z=0$. Blue, red and yellow lines show the median contribution from stellar sources of dust production: AGB stars,
      type II supernovae and type Ia supernovae respectively, with the dashed lines denoting the 84 and 16
      percentiles. The green line shows the contribution from grain growth inside molecular clouds. The black line shows
      the dust destruction rate.}
    \label{fig:rate_stellar}
  \end{center}
\end{figure}

The detailed dust model we have built includes several different dust production and destruction
mechanisms that all contribute to the final dust properties of the galaxies in our model.
Figure~\ref{fig:rate_redshift} shows the mean dust production (or destruction) rate densities as a
function of redshift for galaxies in the (480\,$h^{-1}$Mpc)$^3$ MR simulation. The total dust destruction rate plotted includes destruction from supernovae, star formation, and reheating. We also
plot the star formation rate density for comparison.

We can see that grain growth in molecular clouds dominates the production of dust over the redshift range $z =
0-8$, rapidly increasing towards its peak at $z = 2$. The destruction rate closely follows the dominant grain growth production rate, suggesting that any dust destroyed is rapidly recycled by grain growth. While at very early times \tTwo\ supernovae dominate the production of dust. Thus, at the highest redshifts, the dispersion in the DTM ratio is small, with the dispersion increasing rapidly as grain growth takes over at $z<8$.

If we look at the stellar contributions to the dust content, we see that \tTwo\ supernovae are the dominant stellar production
mechanism across the whole redshift range, peaking at $z\sim2$, closely following the shape of the star formation rate as one would
expect. Dust production (and metal enrichment) from \TOneA\ supernovae is shifted to slightly later times, due to the power-law delay-time distribution (DTD) we assume, which allows $\sim{}52$ per cent of the supernovae to explode $> 400$ Myr after star formation (see \citealt{Yates2013}). Nonetheless, \TOneA\ supernovae never have a
significant impact on the dust production rate. It is worthwhile to note that many other works also suggest that \TOneA\ 
supernovae are unlikely to be the major sources of ISM dust \citep[\eg][]{Nozawa_2011}. 
Production by AGB stars is also negligible at early times, but rises at
late times to rates approaching that of \tTwo{}s.

It is important to note that, although grain growth is the dominant dust formation mechanism at all redshifts below
$z=8$ when averaged over the whole galaxy population, the dust content of individual galaxies can vary enormously.  At
$z=6$, for example, grain growth exceeds stellar dust injection by a factor of 6, but the spread in DTM ratios seen in
Figure~\ref{fig:DTM_fractions} extends over more than a decade.

The variation of the dust production rates with stellar mass is shown in Figure~\ref{fig:rate_stellar} for $z=0$ for star forming galaxies (defined here as galaxies with a specific star formation rate, sSFR\,$>1/3t_{\rm{H}}(z)$, where $t_{\rm{H}}(z)$ is the age of the Universe at redshift z). From this it is clear that there is very little dependence of dust growth and destruction upon galaxy mass. The same holds too at all other redshifts. 

If we compare our dust production rates with the \citetalias{Popping2017} model (their Figures 8 \& 10), this is bound
to be different since the models differ in the grain growth implementation as well as the dust yield tables used
for stellar production mechanisms. But the trends seen in both the models are similar in the sense that grain growth
dominates over all the other production mechanisms at almost all redshifts from z = 0-8. In their model, the median grain growth rate is approximately 3 orders of magnitude higher than any stellar production mechanisms for high stellar mass
($>10^{10}M_{\odot}$) at all redshifts. In our model, the dust production rate from SNII and grain growth is similar
at z $\sim$ 8. Also, we note that the production rates for each of the various sources (SNe-II, AGB stars, and grain growth) are similar between the two models at high mass, whereas they are three to four orders of magnitude greater at low mass in our model compared to \citetalias{Popping2017}. These differences in the dust production rates from different processes are reflected in slight
differences seen in our results for the dust content of galaxies, discussed in the next section.

%% file: content.tex
\section{Results: Dust content of galaxies}
\label{sec:resultsContent}

In this section we compare the predicted dust content of galaxies in our model to observations such as the dust-to-gas
ratio, the stellar-mass -- dust-mass relation and the dust mass function.

\subsection{Dust-to-Gas (DTG) ratio}

\label{sec:resultsDTG}
\begin{figure}
  \begin{center}
    \includegraphics[width=0.48\textwidth]{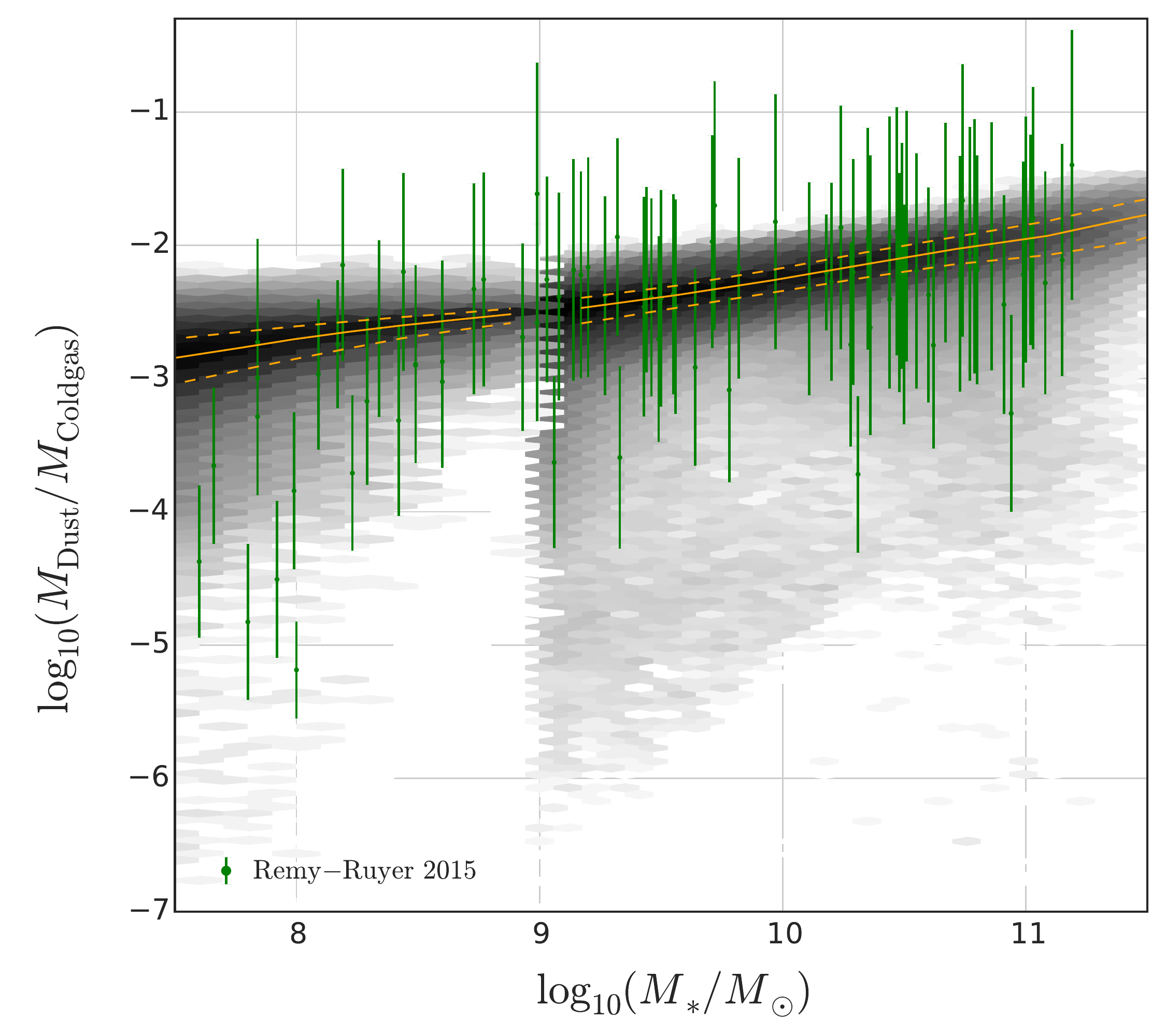}
    \centering
    \caption{The dust-to-gas ratio as a function of stellar mass for z = 0. The orange line shows the median result
      from galaxies in our model, with the dashed lines denoting the 84 and 16 percentiles. Green
      points show the observational constraints from \protect\cite{RR2014}.}
    \label{fig:DTG_stellar}
  \end{center}
\end{figure}

\begin{figure}
  \begin{center}
    \includegraphics[width=0.48\textwidth]{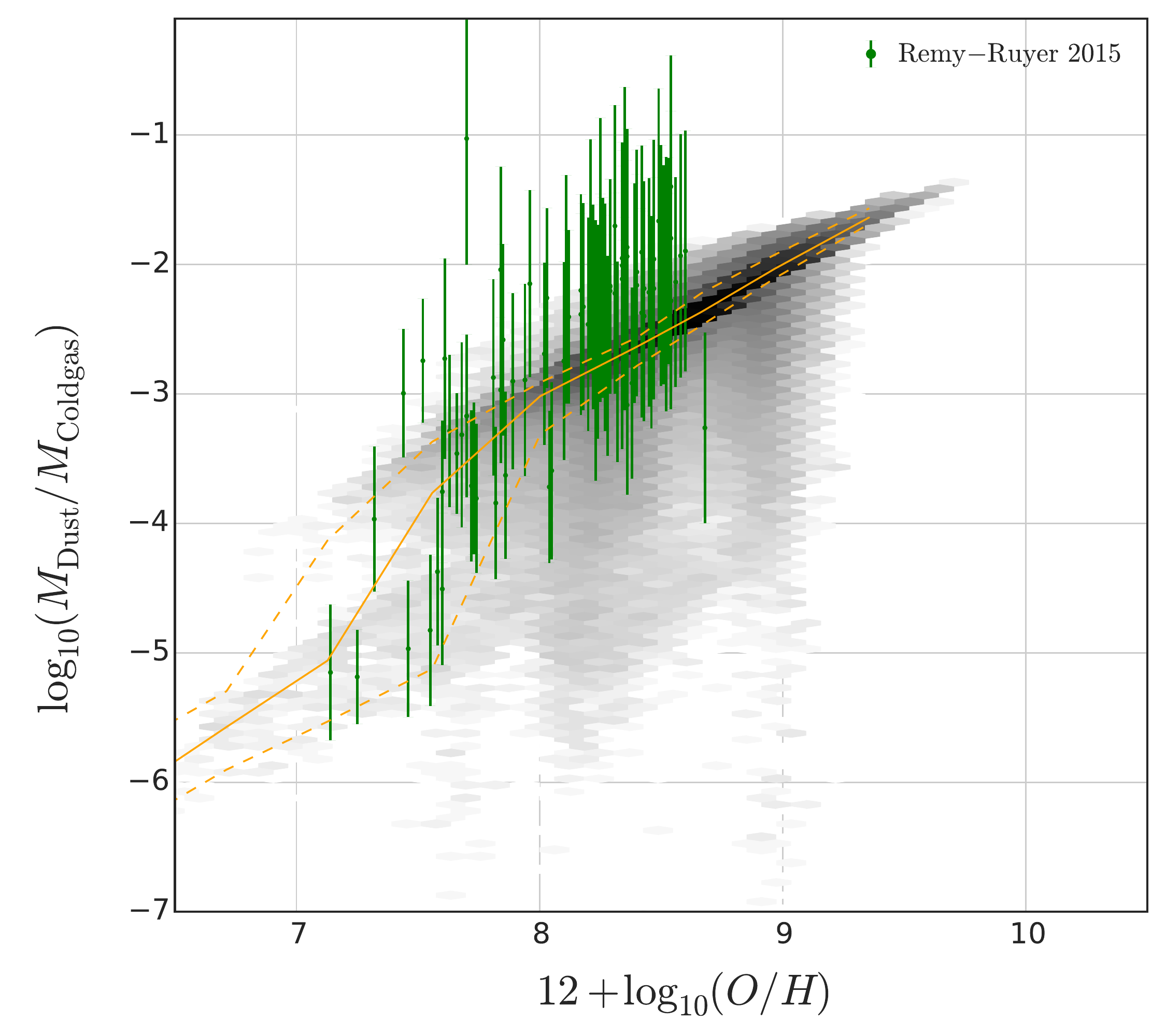}
    \centering
    \caption{The dust-to-gas ratio as a function of gas phase metallicity for z = 0. The orange line 
      shows the median result from galaxies in our model, with the dashed lines denoting the 84 and 
      16 percentiles. Green points show the observational constraints from \protect\cite{RR2014}.}
    \label{fig:DTG_oxygen}
  \end{center}
\end{figure}

We compare the DTG ratio to two
different properties, first, to see how the DTG ratio varies with stellar mass in Figure~\ref{fig:DTG_stellar}, and
secondly how it varies with oxygen abundance, as seen in Figure~\ref{fig:DTG_oxygen}.  Because of the difficulty in
obtaining observational data for comparison, we show only results for $z=0$; at higher redshifts, the DTG ratio exhibits
the same behaviour seen for the DTM ratio in \Fig{DTM_stellar}. 

In Figure~\ref{fig:DTG_stellar}, we compare the DTG ratio of our model versus stellar mass against observations from
\cite{RR2014}. The median value of our model fits the observations well, particularly above stellar masses of
$10^8$M$_\odot$.  Below this value, there may be a downturn in the DTG ratio in the data, that we do not
see. Figure~\ref{fig:DTG_oxygen} shows the same data plotted as a function of oxygen abundance and here we see that the
low DTG ratios are associated with low metal abundance, and that the observations and the model overlap quite well.  The
reason for the discrepancy seen at low masses in Figure~\ref{fig:DTG_stellar} is therefore due to the fact that our
low-mass galaxies mostly have higher oxygen abundance than those in the \cite{RR2014} sample. 

The \citetalias{Popping2017} (their Figure 4 and 3) as well as \cite{Hou2019} (their Figure 4a and 4b) model exhibits a similar trend to our predictions when the DTG ratio is plotted as a function of stellar mass and metallicity respectively. But at all redshifts both the models exhibits a steeper slope, such that their lower mass model galaxies have lower DTG ratios. In case of \cite{McKinnon2016a}, the DTG ratio shows a flat trend with metallicity (their Figure 8) for 12 + log(O/H) > 8 while showing a positive correlation below that. 

\subsection{Dust versus stellar mass}
\label{sec:resultsDustMstar}

\begin{figure*}
  \begin{center}
    \includegraphics[width=\textwidth]{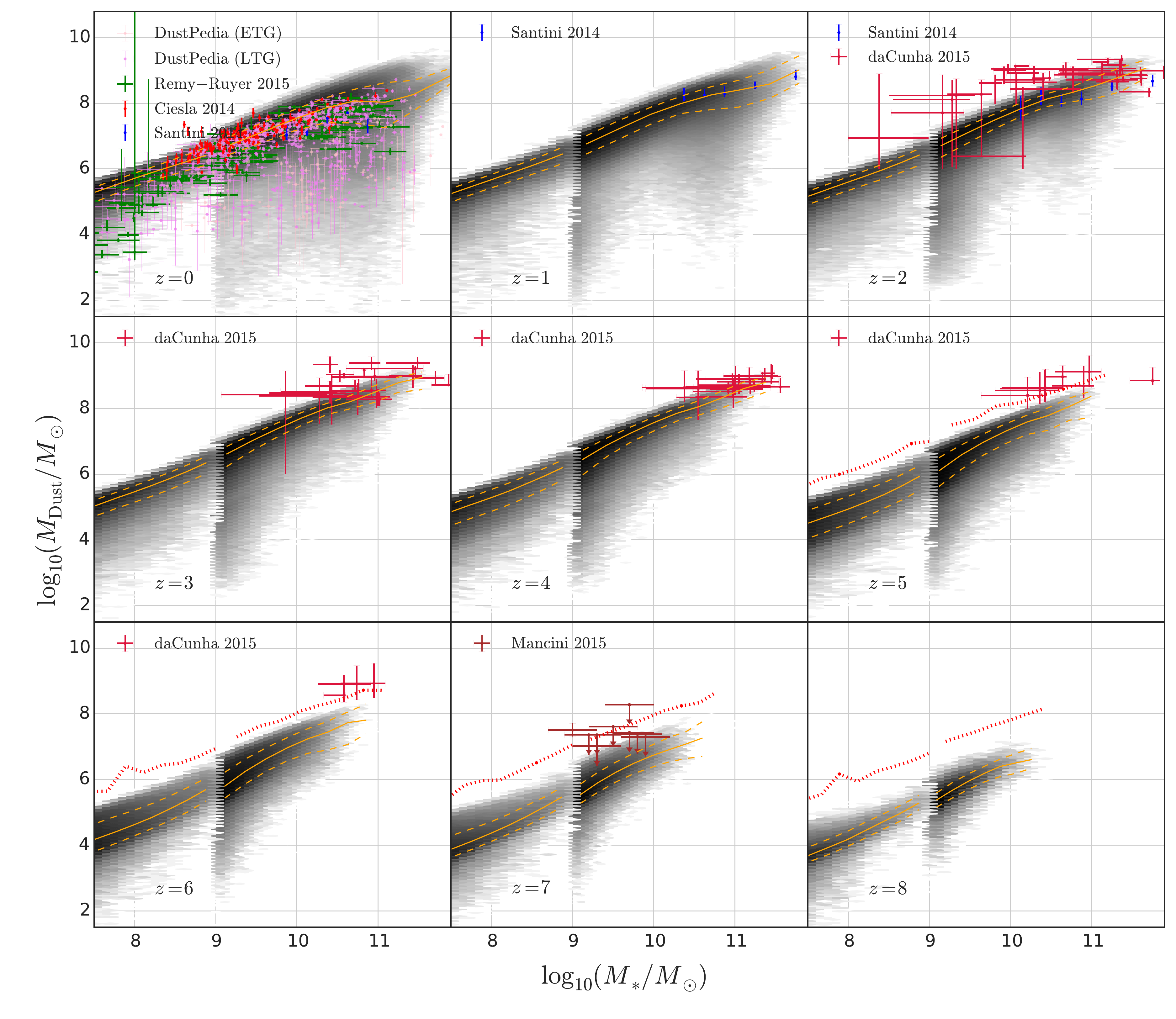}
    \centering
    \caption{The stellar-dust mass relation for redshifts $z=0-8$. The orange line shows the median
      result from galaxies in our model, with the dashed lines denoting the 84 and 16
      percentiles. Pink, violet, red, green, blue, crimson and brown points show the observational constraints from the DustPedia archive \citep[see][separated into ETGs and LTGs]{Davies_2017}, \protect\cite{Ciesla2014}, \protect\cite{RR2014}, 
      \protect\cite{Santini2014}, \protect\cite{daCunha2015} and \protect\cite{Mancini2015} respectively. At
        redshifts 5 and above, the dotted red line shows the maximal dust content that could be predicted by our model,
        assuming saturated grain growth and no dust destruction. }
    \label{fig:stellar_dust_allz}
  \end{center}  
\end{figure*}

The dust mass versus stellar mass relation is shown in \Fig{stellar_dust_allz}.
The evolution in dust masses mimics that shown in \Fig{DTM_stellar} for the DTM ratio. At $z=0$ most of the
galaxies have saturated dust growth on grains.  This persists up to $z=4$, after which there is a gradual transition
down to the levels expected for dust injection from stellar sources.

The stellar-dust mass parameter space is one where we have observational constraints across a very large range of
redshifts. The coloured points in \Fig{stellar_dust_allz} represent observations from a number of different studies \citep[DustPedia
  collaboration][]{Davies_2017,Ciesla2014,daCunha2015,Mancini2015,RR2014,Santini2014}. The DustPedia data combine the
Herschel/Planck observations with that from other sources of data, and provide observations at numerous wavelengths
across the spectral energy distribution. The dust masses are fitted using CIGALE assuming either the dust model from \cite{Draine2014} or their own called THEMIS (see \citealt{Davies_2017}). We use the dust masses fitted by the
former model, since the latter has a lower normalisation at $z=0$ compared to our dust masses. The
\cite{Ciesla2014} data uses the Herschel Reference Survey \citep{Boselli2010}, where the dust masses are obtained using
the SED templates described in \cite{Draine2007b}. \cite{daCunha2015} derives dust masses from a sample of sub-mm
galaxies in the Alma LESS survey using the SED fitting techniques described in \cite{daCunha2008}. Some of the galaxies
in the sample only have photometric redshifts and thus the redshift is kept as a free parameter in their fitting
technique. \cite{Mancini2015} uses ALMA and PdBI observations with upper limits on the dust continuum emission. They
derive the stellar masses using the mean relation between the UV magnitude and the dust mass assuming T$_d$ = 35K and
$\beta$ = 1.5. \cite{Santini2014} uses galaxies in the GOODS-S and GOODS-N field as well as the COSMOS field which have
FIR observations carried out using \textit{Herschel}. They also use the SED templates of \cite{Draine2007b} as a
description for their dust masses.

The first thing to note is that there is a significant offset in normalisation between the different
observational data sets at $z=0$. Thus we see that, while the median dust content predicted by our
model is consistent with the LTGs from DustPedia and \citet{Ciesla2014} data, it lies well above that of 
\cite{RR2014} and \cite{Santini2014}. This reflects the different observational biases and systematic uncertainties 
in the estimation of dust content. For example, the \cite{RR2014} sample contains some massive AGN-host galaxies which are presumably older and have low gas fractions, leading to smaller dust masses. Also a part of their sample \citep[DGS,][]{Madden2013} was chosen to study low-metallicity environments and hence exhibit smaller dust masses. 

Although the median dust level is acceptable, it would appear that we have many galaxies,
particularly at masses above about 10$^{10}$\,\Msun, whose dust content is significantly higher than those seen in the observational samples considered here.
This could come about in one of three ways: too much cold gas; too
high a metallicity in the cold gas; too high a dust-to-metal (DTM) ratio.  The cold gas content of
galaxies in the \citetalias{Henriques2015} model was considered in \citet{Martindale2017} and while the
H{\sc i} mass function was in good agreement with the observations, the gas-to-stellar mass ratio
is, if anything, slightly too low (although the selection functions for the H{\sc i} surveys are
hard to reproduce).  Similarly, \citet{Yates2013} showed that the oxygen abundance of cold gas in our
model is in good agreement with observations from SDSS.  Finally, \Sec{resultsDTM} of this paper
shows that the DTM ratio is in good agreement with that of \citet{RR2014}.  It is thus slightly
perplexing that we seem to have these galaxies with excessive dust.  We note that in our model, we
have ignored possible dust destruction due to the effects of cosmic rays, photoevaporation or AGN
activity that start to play a major role in high mass galaxies.

There is also a significant spread in observed dust masses to lower values at high stellar masses at $z=0$ due to the
presence of elliptical early-type galaxies (ETGs) with low molecular gas content.  We predict many such galaxies in our
model (see also Figure~\ref{fig:DTM_stellar_age}) but in a lower proportion than in the DustPedia data set -- it is
unclear to what extent this is an observational selection effect.

At higher redshifts, up to $z=4$, the upper locus of our dust masses lies, if anything, slightly below the observations,
and at $z=5,6$ and 7 it is well below.  We note, however, that almost all of the Mancini data are upper limits, and that the
\cite{daCunha2015} data are ALMA observations of sub-mm galaxies which are some of the brightest star-forming galaxies
at that particular redshift, hence a population biased towards more dust-rich systems. The dotted red lines in
Figure~\ref{fig:stellar_dust_allz} show the saturation value (as discussed in Section~\ref{sec:resultsDTM}). To reproduce any observations lying above this would require either a higher cold gas content, or a higher metallicity (i.e.~earlier
enrichment), or too high a destruction rate in the semi-analytic model. It is worthwhile to note that the dust destruction efficiency adopted in this study is based on calculations for multiphase ISM in the solar environment, hence one could imagine the ISM having different properties at $z>5$, thus also changing the dust destruction rates.

\citetalias{Popping2017} also found mixed success in matching observations of the stellar mass -- dust mass relation (their
Figure 2) in both local and high-redshift galaxies. At $z = 0$, their median relation lies below the observations of \citet{Ciesla2014}, but follows the trend seen by \citet{RR2014} at low mass, where they reproduce a steep stellar mass -- dust mass relation. This is chiefly due to the longer accretion timescales they assume at low molecular gas densities, which can reach around 1 Gyr (see their Figure~1), compared to values closer to 10 Myr for this work (see Figure~\ref{fig:App stell tacc}). They have galaxy masses up to $3\times 10^{11}$\,\Msun\ at all redshifts up to $z=9$, finding a median dust-to-stellar mass relation with a steeper slope than our results, thus providing a better match to the high redshift observations than we do.
We note that these differences in our results are driven by the strong molecular-gas dependence in their empirical $\tau\sub{acc}$ prescription, which is in turn driven by the enhanced star-formation efficiency they assume at $\Sigma\sub{H2,crit} > 70 \Msun/\tn{pc}^{-2}$ (their Equation~1); our model assumes much smaller variations in the properties of molecular clouds in galaxies of different surface densities.  

\subsection{Dust mass function}
\label{sec:dustMF}
\begin{figure}
  \begin{center}
  	\centering
    \includegraphics[width=0.4\textwidth]{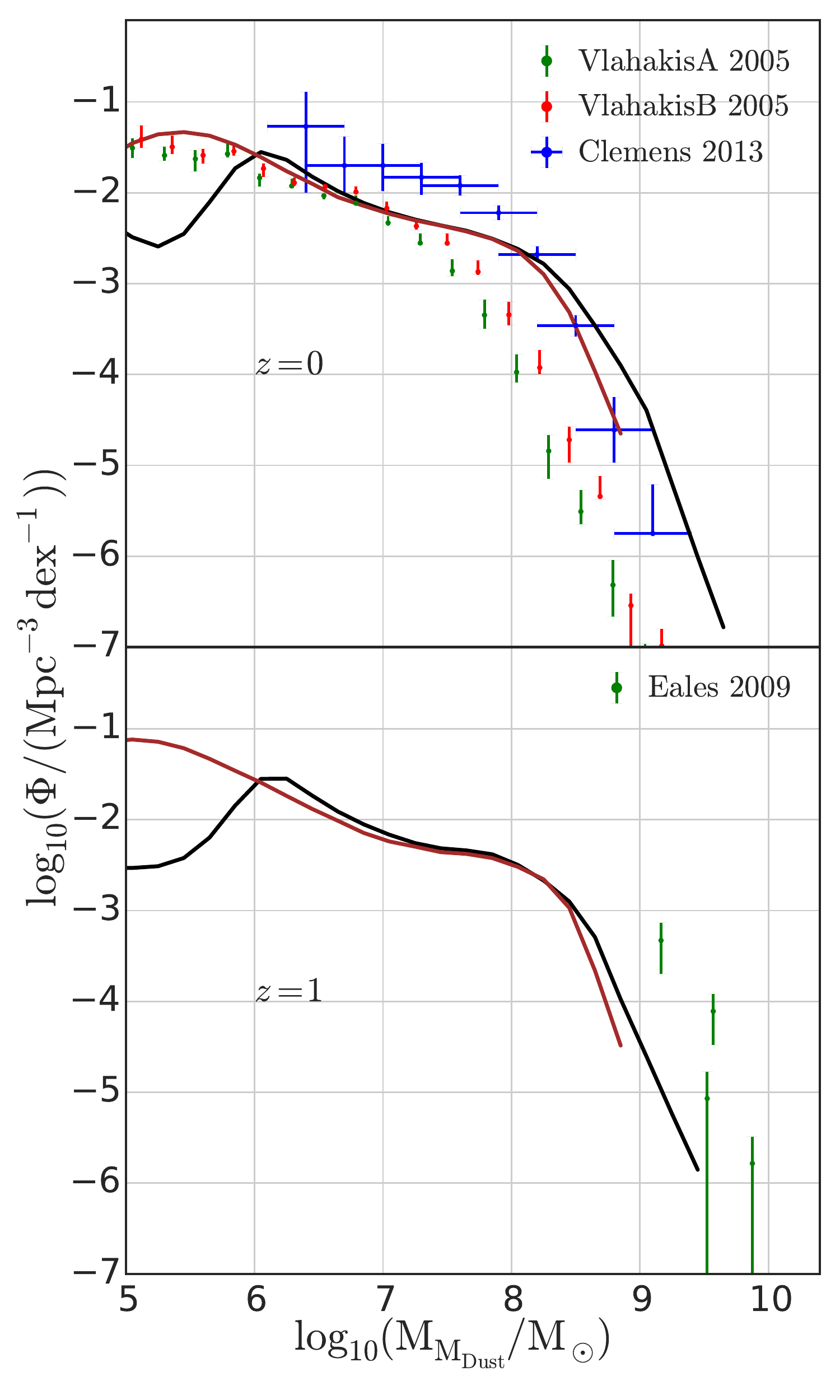}
    \caption{The Dust Mass Function (DMF) for redshifts $z = 0$ and $1$. The black line shows the
      prediction of our model using the underlying dark matter Millennium simulation, and the red
      line for Millennium-II. Observations are taken from \protect\cite{Vlahakis2005} and
      \protect\cite{Clemens2013} at $z=0$, and from \protect\cite{Eales_2009} for $z=1$. }
    \label{fig:DMF}
  \end{center}
\end{figure}

Figure~\ref{fig:DMF} shows the dust mass function at $z=0$. The red line shows the
results of the Millennium-II simulation, and the black line the Millennium simulation. We compare
with observations from \cite{Vlahakis2005}, \cite{Clemens2013} and \cite{Eales_2009}. \cite{Vlahakis2005}
derived the local sub-mm luminosity and dust mass functions using the SCUBA (the Submillimetre
Common-User Bolometer Array) Local Universe Galaxy Survey (SLUGS) and the IRAS Point Source Catalog
Redshift Survey (PSCz). They fit two component grey bodies to their SEDs with emissivity index
$\beta = 2$ and dust temperature in the range 17-24\,K.  The `A' sample determines dust masses using
a dust temperature obtained from isothermal SED fitting, and the `B' dust mass function has been
calculated using a dust temperature of 20\,K.  \cite{Clemens2013} combined Herschel data with
Wide-field Infrared Survey Explorer (WISE), Spitzer and Infrared Astronomical Satellite (IRAS)
observations to investigate the properties of a flux-limited sample of local star-forming
galaxies. They fit their SEDs with modified blackbody spectra using $\beta\simeq 2$ and
dust temperatures in the range 10-25\,K. \cite{Eales_2009} uses data obtained from the 
Balloon-borne Large Aperture Submillimeter Telescope (BLAST), using the greybody relation assuming 
a dust temperature of 20K.

We find that the model provides a good fit to the \cite{Vlahakis2005} observations at low and intermediate dust masses,
but under-predicts the number density when compared with \cite{Clemens2013} at the same mass range. The knee of the mass
function is at a lower mass in \cite{Vlahakis2005} compared to our model output, while in \cite{Clemens2013} it roughly
coincides with our model. At the high mass end, our predicted number densities are higher than both the observational
data sets.  This result is consistent with that of the previous section, that we over-predict the dust content of many
massive galaxies at $z=0$ in our model. On comparing our model predictions to the \cite{Eales_2009} data for $z=1$, we
instead appear to slightly under-predict the dust mass function at high masses. It is worthwhile to note that this is a 
general feature seen in other models of galaxy formation tracking dust growth (\eg\ \citealt{McKinnon2016}, 
\citetalias{Popping2017}).

%% file: conc.tex
\section{Conclusions}\label{sec:conc}

We have run a modified version of the {\sc L-Galaxies} semi-analytic model which includes a prescription of dust
modelling on the full Millennium and Millennium-II trees.  By combining both the Millennium simulations we are able to
make use of both the higher volume in order to find rarer objects, but also the higher mass resolution of Millennium-II
to probe lower mass galaxies. Our conclusions are as follows:

\begin{enumerate}

\item Our grain growth model follows that of previous work, as described in \citet{Popping2017}, but following
separately the dust content in molecular clouds and the inter-cloud medium.  We find that, in regimes where $\tau_{\rm{exch}}\gg\tau_{\rm{acc}}$ as well as for low values of $\mu$, this
can have a significant impact upon the dust growth rate (\Fig{fc_fd}).

\item The dust-to-metal (DTM) ratio (\Fig{DTM_stellar}) shows an evolution from low to high ratios, the former
  corresponding to dust injection from \tTwo\ supernovae, and the other to maximal, saturated dust production occurring
  via dust growth on grains. The latter dominates at redshifts below $z\approx 4$. A significantly populated transition
  region is seen at $z=6$.

\item By colouring with age (\Fig{DTM_stellar_age}) we show that this is the primary driver of the movement from low to
  high DTM ratio.

\item When plotted as a function of gas-phase metallicity, we find a reasonable fit to the observations at all redshifts
  (\Fig{DTM_oxygen}).

\item We present a fitting relation for the DTM ratio, dependent on the metallicity and mass-weighted age of the galaxy
  stellar population. That provides a good fit to the model at both low and high redshift, but with some scatter at
  intermediate redshifts due to the varied growth histories of galaxies (\Eq{DTMfit} and \Fig{DTM_fit}).

\item Grain growth is the dominant dust production mechanism at all redshifts below $z=8$ (\Fig{rate_redshift}).  Dust
  destruction rate closely follows the grain growth production rate, suggesting prompt recycling of any dust content.
  We note, however, that \Fig{DTM_stellar} shows that by $z=6$ only half of galaxies lie on the upper locus of
  DTM-ratio.  Thus the detailed history of galaxy formation is important for determining the dust content of any
  individual galaxy.

\item The dust growth rates show little dependence on galaxy mass (\Fig{rate_stellar}).

\item We find a good fit to the shape and normalisation of the dust-to-gas ratio at $z=0$ when plotted as a function of
  both stellar mass (\Fig{DTG_stellar}) and oxygen abundance (\Fig{DTG_oxygen}).

\item We find a reasonable fit to the shape and normalisation of the observations in the stellar-dust mass plot
  (\Fig{stellar_dust_allz}) over a wide range of redshifts, $z = 0-4$.  We have an excess of very dusty, massive
  galaxies at $z=0$, perhaps due to a lack of destruction mechanisms. we fail to predict the dustiest galaxies at $z>5$,
  which hints that our dust growth rate may be too slow, or the destruction rate too high; however, we note that the
  interpretation of the observations are very uncertain at these redshifts.

\item There is a good agreement between the predicted $z=0$ dust mass function at the intermediate and low dust masses
  with observations; however we over predict the number density of galaxies at the highest dust masses
  (Figure~\ref{fig:DMF}). This again suggests that we may have too much dust in the most massive galaxies.

\end{enumerate}

The model that we have presented here is deficient in at least 2 respects. Firstly, it assumes that dust is instantly
destroyed in the hot (coronal) phase of the interstellar medium. Secondly, we ignore the effect of dust on the physics of
galaxy formation: the formation of molecules on grains, and the coupling to radiative feedback, for example.  This will
be investigated in future work.

It seems evident from our work that, at sufficiently high redshift, there will be a transition from high (saturated dust
growth) to much lower (primarily \tTwo\ supernovae) dust-to-metal ratios.  The precise redshift at which this happens
depends upon uncertain grain growth and destruction time-scales.  Nonetheless, it is important to appreciate that there
will be a wide variety of DTM ratios in galaxies at high redshift.  The situation will become much clearer over the next
few years with deep extragalactic surveys such as those proposed by {\it EUCLID, WFIRST} and {\it JWST} and follow-up
with ground-based observations from facilities such as {\it ALMA}.

%% file: acc.tex
\section{Accretion Timescale}\label{sec: acc_scale}

\begin{figure*}
  \begin{center}
    \includegraphics[width=\textwidth]{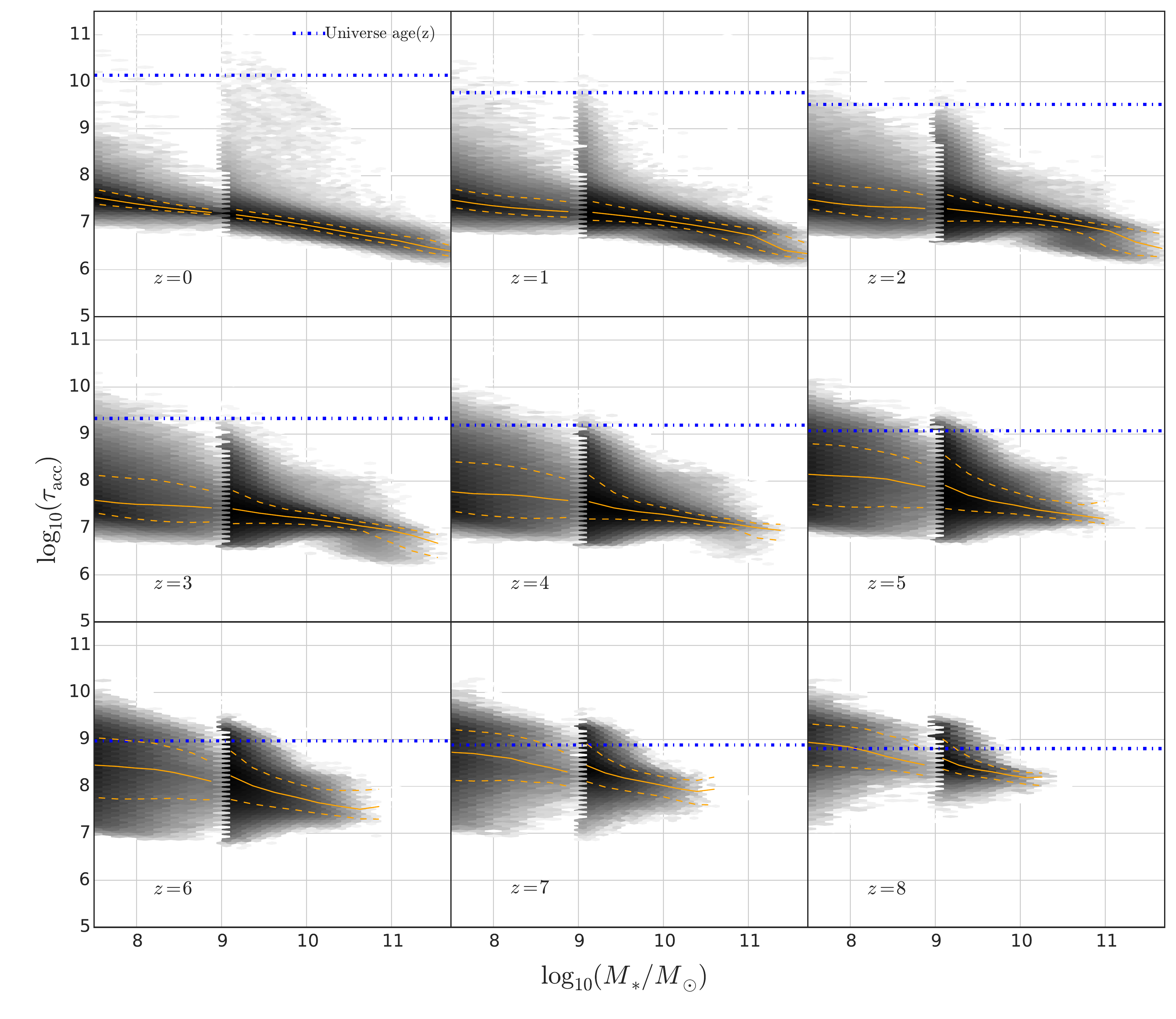}
    \centering
    \caption{The accretion timescale-stellar mass relation for redshifts $z=0-8$. The orange lines show the median
      result from galaxies in our model and the 1-sigma scatter. The age of the Universe at that particular redshift is
      shown as the dot-dashed blue line. }
    \label{fig:App stell tacc}
  \end{center}
\end{figure*}

\begin{figure*}
  \begin{center}
    \includegraphics[width=\textwidth]{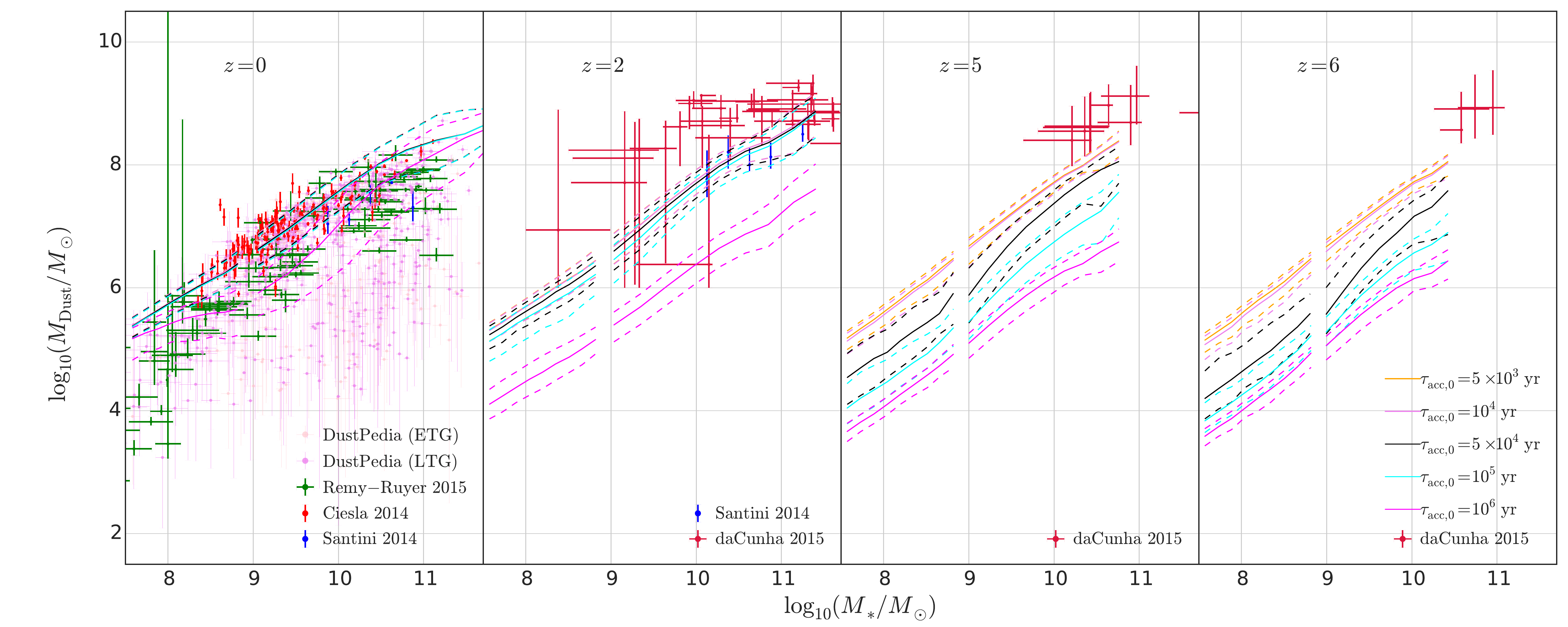}
    \centering
    \caption{The median dust-stellar relation in our model for different values of $\tau_{\rm{acc,0}}$ at $z = 0$, 2, 5
      and 6. The solid line shows the median relation while the dashed lines denotes the 84 and 16 percentiles. The
      observational constraints from the DustPedia archive \protect\citep[see][separated into ETGs and
        LTGs]{Davies_2017}, \protect\cite{Ciesla2014}, \protect\cite{RR2014}, \protect\cite{Santini2014} and
      \protect\cite{daCunha2015} respectively have been plotted for comparison.}
    \label{fig:App diff tacc}
  \end{center}
\end{figure*}

\begin{figure*}
	\begin{center}
		\centering
		\includegraphics[width=\textwidth]{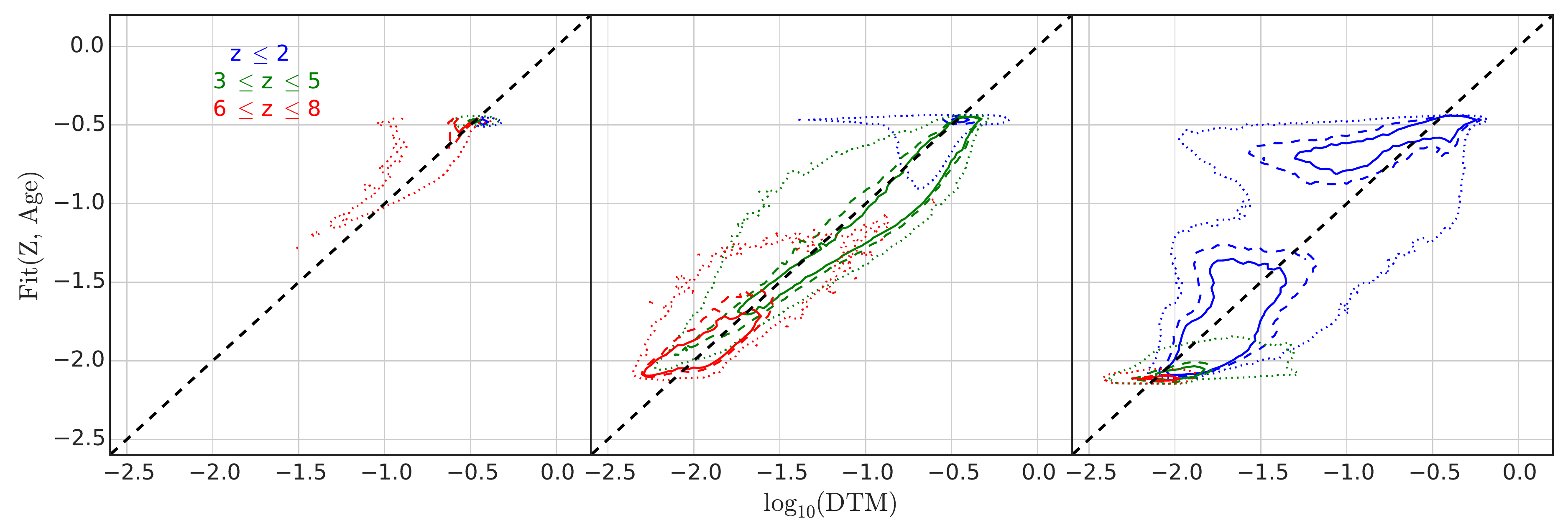}
		\caption{The DTM fitting function in Equation~\ref{eq:DTMfit} is plotted against the DTM ratio
			from the model, similar to Figure~\ref{fig:DTM_fit}. From left to right DTM values generated by running our model for $\tau_{\rm{acc,0}}$ values of 10$^4$, 10$^5$ and 10$^6$ yr respectively are plotted against the fit function.}	
		\label{fig:App diff tacc fits}
	\end{center}
\end{figure*}

Here we will discuss how the accretion timescale varies with redshift as well the impact of choosing a different
$\tau_{\rm{acc,0}}$ on our dust model.

Figure~\ref{fig:App stell tacc} shows the distribution of $\tau_{\rm{acc}}$ plotted against stellar mass for $z
=0-8$. The age of the Universe at each redshift is also plotted for comparison. The median value of $\tau_{\rm{acc}}$
moves towards lower values as we move to lower redshifts due to the increase in the DTG ratio (see
equation~\ref{eq:tacc}). Note that there are a lot of galaxies at high redshift ($z \geq 6$) that have $\tau_{\rm{acc}}$
values similar to the age of the Universe at that particular redshift -- this is also the reason for very low values of
DTG or DTM ratio, with comparable or higher values of the stellar production rate compared to grain growth.  As we move
towards lower redshift, most $\tau_{\rm{acc}}$ values start to dip beneath the age of the Universe and at $z \le 2$ the
median values are 3 to 4 orders of magnitude less than the age of the Universe.  Thus the choice of $\tau_{\rm{acc,0}}$
has a negligible effect at low redshifts but can be be quite significant in determining the galaxy dust mass at high
redshifts.

To see the effect of modifying the value of $\tau_{\rm{acc,0}}$ on the galaxy dust mass we consider values ranging from
$5\times10^{3}-10^{6}$\,yr.  The median dust-stellar mass relation for $z = 0$, 2, 5 and 6 with these accretion
timescales are shown in Figure~\ref{fig:App diff tacc}. The dust-stellar mass relation at $z = 0$ is not drastically
affected by changes in $\tau_{\rm{acc,0}}$, except for $\tau_{\rm{acc,0}} = 10^{6}$\,yr where the median is about
0.5\,dex lower than the other median values at intermediate stellar masses -- this is because the dust growth timescale
becomes comparable to the destruction timescale. Similarly, at $z = 2$ the DTM ratio for $\tau_{\rm{acc,0}} =
10^{6}$\,yr has decreased by more than an order of magnitude. At $z = 5$ and 6 the changes are more visible with a
spread in DTM ratios becoming apparent as $\tau_{\rm{acc,0}}$ is varied. The main point to take away from this is that grain
growth requires time to act, and that timescale depends on the value of $\tau_{\rm{acc,0}}$.

We also compare how our fitting function, Equation~\ref{eq:DTMfit} performs for different values of $\tau_{\rm{acc,0}}$ in Figure~\ref{fig:App diff tacc fits}. For this we ran our model with $\tau_{\rm{acc,0}}$ values of 10$^4$, 10$^5$ and 10$^6$ yr, and obtain the expected DTM ratio using the corresponding $\tau_{\rm{acc,0}}$ values in Equation~\ref{eq:DTMfit}. We see that the fit does a good job for $\tau_{\rm{acc,0}}=10^4$\,yr where we expect DTM ratios to be near saturation, while for the higher $\tau_{\rm{acc,0}}$ values we see considerable scatter in the fit. This scatter for $\tau_{\rm{acc,0}}=10^6$\,yr directly follows from our previous discussion of the grain growth timescales. This has led to a bimodal distribution at $z\le2$, with the grain growth dominated population at the top and the stellar injection dominated ones at the bottom. The sharp cut-off in the bottom population is an artifact of our fitting function, as it can not have values less than $\mathcal{D}_0$.